\documentclass[aps,pre,amsfonts,amssymb,amsmath,twocolumn,superscriptaddress]{revtex4-2}
\usepackage{graphicx}     
\usepackage{dcolumn}      
\usepackage{bm}           
\usepackage[T1]{fontenc}
\usepackage{color}
\usepackage[breaklinks=true]{hyperref}
\hypersetup{colorlinks=true,linkcolor=blue,filecolor=blue,urlcolor=blue,citecolor=blue}
\usepackage{tikz}
\usetikzlibrary{plotmarks}
\newcommand\marksymbol[2]{\tikz[#2,scale=1.2]\pgfuseplotmark{#1};}
\usetikzlibrary{shapes}
\newcommand\mystar{\tikz{\node[draw=cyan,star, fill=cyan,star point height=.6em,scale=0.3]{};}}
\begin{document}
\title{Phase diagram in a one-dimensional civil disorder model}
\author{Ignacio Ormaz{\'a}bal}
\email[Corresponding author: ]{iormazabal@udec.cl}
\affiliation{Departamento de F{\'i}sica, Universidad de Concepci{\'o}n, Concepci{\'o}n, Chile}

\author{Felipe Urbina}
\affiliation{Centro de Investigaci{\'o}n DAiTA Lab, Facultad de Estudios Interdisciplinarios, Universidad Mayor, Santiago, Chile}

\author{F{\'e}lix A. Borotto}
\author{Hern{\'a}n F. Astudillo}
\affiliation{Departamento de F{\'i}sica, Universidad de Concepci{\'o}n, Concepci{\'o}n, Chile}
\date{\today}
\begin{abstract}

Epstein's model for a civil disorder is an agent-based model that simulates a social protest process where the central authority uses the police force to dissuade it. The interactions of police officers and citizens produce dynamics that do not yet have any analysis from the sociophysics approach. We present numerical simulations to characterize the properties of the one-dimensional civil disorder model on stationary-state. To do this, we consider interactions on a Moore neighborhood and a random neighborhood with two different visions. We introduce a Potts-like energy function and construct the phase diagram using the agent state concentration. We find order-disorder phases and reveal the principle of minimum grievance as the underlying principle of the model's dynamics. Besides, we identify when the system can reach stable or an instability conditions based on the agents' interactions. Finally, we identified the most relevant role of the police based on their capacity to dissuade a protest and their effect on facilitating a stable scenario. 

\end{abstract}
\maketitle
\section{Introduction}

In recent years, the sociophysics or statistical physics of social dynamics has described different social phenomena as collective effects of the interaction between individuals \cite{castellano2009a,galam2012,psen2014}. In particular, the study of opinion dynamics has generated various models describing consensus, agreement, or uniformity using tools from statistical physics \cite{Galam2008,Redner2019,Jedrezejewski2019,Sznajdweron2021}. Recent efforts aim at describing these models considering a diversity of individual traits of a population and size of group discussion \cite{Galam2020}, or their multi-state variations, such as, for example, the majority voting model \cite{li2016}, the multi-state voter model \cite{vazquez2019}, multichoice opinion dynamics models \cite{bancerowski2019}, or the multi-state noisy q-voter model \cite{nowak2021}.

In parallel, social scientists have used agent-based models to reproduce emerging social phenomena \cite{epstein1996, epstein2006}, such as the Schelling model of urban segregation \cite{schelling1969} and the Axelrod model of cultural dissemination \cite{axelrod1997}. These models have attracted the attention of physicists, who have described the Schelling model as interacting physical particles \cite{vinkovic2006} and as an Ising-like model \cite{stauffer2007}. Furthermore, they have characterized the static and dynamic properties in one and two dimensions \cite{dallasta2008} and their different behaviors using a phase diagram \cite{gauvin2009}. They have recently used a similar energy function to characterize the Schelling and Sakoda models \cite{Pollicott2001,goles2011,medina2017}. On the other hand, physicists described the Axelrod model in two dimensions showing order-disorder phase transitions \cite{castellano2000}. Then, they described the one-dimensional Axelrod model as a starting point for its description in more complex topologies \cite{vilone2002}. In addition, they described the role of dimensionality on the order-disorder phase transitions \cite{klemm2003} and the stability model using Lyapunov functions \cite{klemm2005, kuperman2006}. In this context, Epstein presented an agent-based model to describe the social dynamics of protests and rebellions through recognizable macroscopic phenomena \cite{Epstein2002}. This model simulates a social protest process where the central authority uses police force to dissuade a protest.

For a generalized rebellion, Epstein presents five study cases and reports statistical regularities observed in the punctuated equilibrium dynamics, opening new questions about how the civil disorder dynamics work. Different scientists have modified this model to describe other social conflicts such as workers' protests \cite{kim2011}, the spread of criminal activity \cite{fonoberova2012}, or civil war cases between ethnic groups \cite{bhavnani2012}. In addition, some variations include legitimacy with endogenous feedback \cite{lemos2016} or the influence of the distribution of money on the dynamics \cite{ormazabal2017}. Despite these modifications, nobody characterized this agent-based model using concepts and tools of statistical physics in its original form.

This paper aims to characterize the one-dimensional civil disorder model on stationary-state as a first approach to studying this model in other dimensions or topologies. To do this, we perform numerical simulations of the Epstein model with and without police officers and use two visions to define interactions in a Moore neighborhood and a random neighborhood. We introduced two macroscopic quantities and built the phase diagram to identify different behaviors. On the one side, we define a Potts-like energy function to deduce a guiding principle to understand civil disorder dynamics. For the Schelling and Sakoda model \cite{goles2011,medina2017}, this energy-like function allows identifying a minimization principle to understand the spatial segregation patterns as efficient or inefficient. In the case of the dynamics of a factory workers' protest \cite{galam1982}, a function similar to the Ising model's free energy allows identifying the steady-state of the system and describing two phases based on the {\it Principle of Minimum Dissatisfaction}. On the other side, we used the concentration of agents to identify and characterize phase changes. The concentration allows describing phase transitions in the q-voter model with two types of stochasticity \cite{nyczka2012} and the multi-state noisy q-voter model \cite{nowak2021}. Besides, this macroscopic quantity is helpful to building phase diagrams. In the Schelling and Axelrod model \cite{gauvin2009,castellano2000}, this strategy allows identifying domain boundaries of the different qualitative behaviors. Hence, in this work, the agents' concentration and the phase diagram are crucial elements to characterize the original Epstein model and allow us to describe order-disorder transitions not reported in other works.

When considering the system without police officers, we identify transitions from two orders with a majority phase, a disordered phase, and a consensus phase. Furthermore, we identify when the system can reach stability or an instability conditions based on the agent's interactions. Besides, we reveal the {\it Principle of Minimum Grievance}, the underlying principle of the model's dynamics. On the other hand, in the system with police officers, we study the effects of police officers' concentrations in different scenarios generated by the kind of neighborhoods and vision. We find the same order-disorder transition, but now we observe six ordered phases with a majority, one disordered phase, and the consensus phase. With the global quantities that we introduced, we can determine the role of police officers to dissuade a social protest. Finally, we identify stability and instability conditions of the system dynamics, and we show the energetic cost of using the police force to facilitate a stable scenario. These results from the perspective of sociophysics yield new qualitative elements and contribute to the future to study the dynamics of this model in other dimensions and topologies to approach the complexity of the dynamics of social protest.

The paper is organized as follows: In Sec. \ref{model}, we introduce Epstein's model for a civil disorder and the global quantities used to describe the model's behavior to reach the stationary state. The simulations for the model without and with police officers and their respective phase diagrams are presented in Sec. \ref{diagrams}. The discussion of our results and concluding remarks are in Sec. \ref{remarks}.

\section{The model and global quantities}\label{model}
\subsection{The Epstein Model}

The civil disorder model has two agents: citizens and police officers. Citizens can be active when they participate in social protest, passive when they do not participate, or jailed when the police officers catch them. Citizens can switch from one state to another depending on their neighborhood, local parameters, and the global parameters of the system. On the other hand, a police officer agent represents the central authority's force. They are responsible for deterring a protest by capturing the active agents in their neighborhood. The neighborhood for all agents can be a von Neumann neighborhood used by Epstein \cite{Epstein2002} or a Moore neighborhood as in other works \cite{fonoberova2012,ormazabal2017}.

The system's dynamics emerge by relating the legitimacy of the authority and the grievance of the population, i.e., it depends on the relationship between the global parameters and the agents' parameters. The global parameters are the same for all agents: legitimacy $L$, a state change threshold $T$, the maximum jail term $J_{\text{max}}$, and the vision $v$. The original model's vision determines the neighborhood's size, similarly to the rule radius in cellular automata \cite{boccara2010} and range in other opinion models \cite{Castellano2011,roy2014}. On the other hand, the local agent parameters are hardship $H$ and risk aversion $R$. Both parameters are random values between zero and one uniformly distributed among all agents.

\begin{figure}
    \includegraphics[width=0.35\textwidth]{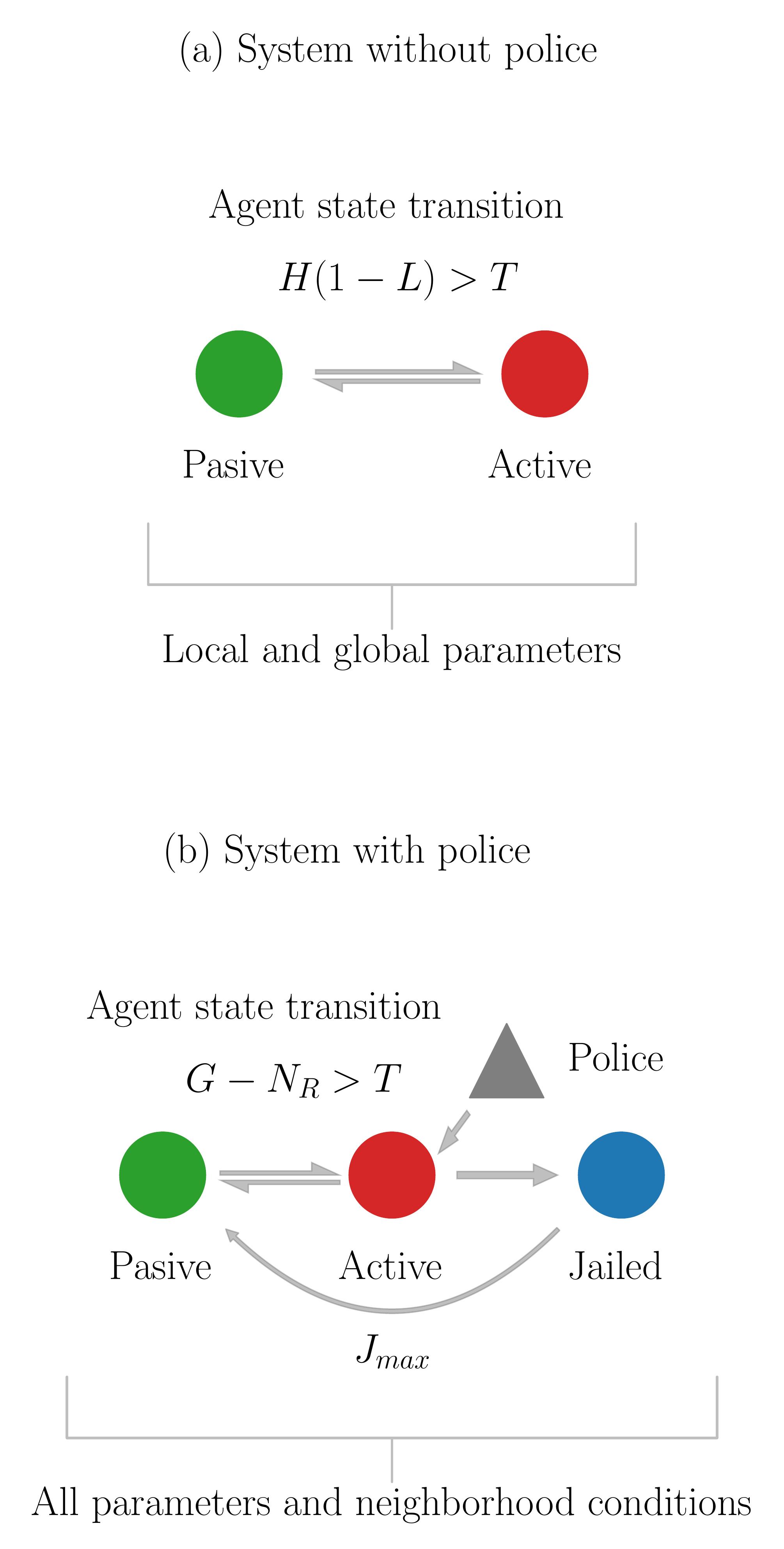}
    \caption{\label{fig_1}Schematic visualization of agents state changes in the Epstein's model for a civil disorder. For the system without police officers, the citizen agents can switch between two states depending on their local and global parameters. For the system with police officers, the agents can change between three states depending on their local and global parameters and neighborhood conditions. The switch from the active to jailed state is a product of the interaction with the police officers' agents.}
\end{figure}

The rules that determine the agents' actions are as follows:

{\it (1) State change rule}: each agent will decide whether or not to join the protest, evaluating the equation $G-N_{R}>T$, where $G=H(1-L)$ symbolizes the grievance and $N_{R}=RP$ the net risk. The arrest probability equation $P=1-\exp[-k(C/A)_{v}]$ depends on the active agents and police officers ratio in the neighborhood defined by the vision. Hence when any agents evaluate if they switch their state, consider all active agents and police officers in their neighborhood. Therefore, for a fixed number of police officers, the agent's arrest probability falls the more active agents there are. Notice that $A$ is always at least one because the agent always counts himself as active when computing $P$. The value of $k$ is $2.3$ for ensuring plausible $P$ values, as reported by Epstein \cite{Epstein2002}. In the complete form of the state switch equation,

    \begin{equation}
        H(1-L) - R(1-\exp[-k(C/A)_{v}]) > T,\label{eq1}
    \end{equation}
    
\noindent we notice that the first element on the left depends on a combination of local and global values, and the other depends on the neighborhood conditions. In this way, when the difference of the agents' state variables exceeds the threshold, they switch from passive to active; otherwise, they remain passive agents.

{\it (2) Capture rule}: police officers randomly capture an active agent from their neighborhood. If there are no active agents, they do nothing. A jailed agent stops participating in the dynamics according to the jail parameter assigned value randomly, with values between zero and the maximum determined at the beginning of the simulation. We used $30$ time steps as a maximum jail term, the same value used by Epstein \cite{Epstein2002}. When jailed agents finish their sentences, they return to the model dynamics as passive agents.

{\it (3) Movement rule}: each agent will move to an empty space at random within their neighborhood.

After setting global and local parameters of the model, we placed all the agents in random positions in the lattice to start the simulations. At each time step, all agents evaluate the dynamics rules asynchronously \cite{Epstein2002}. We show a schematic visualization of the changes of the agents' state changes due to the interaction rules in Fig. \ref{fig_1}. 

\subsection{Global quantities}

In order to characterize the model, we labeled each agent with the variable $\alpha$, which can take values between one and four to represent an active agent, passive, jailed, or police officer. Then, we have defined the following quantities.\\

{\it (a) Concentration of agents}. To see the predominant state in the system and study its macroscopic behavior as a function of the global parameters, we define:

\begin{equation}
    C_{\alpha} = \frac{N_{\alpha}}{N},\label{eq2}
\end{equation}

\noindent where $N_{\alpha}$ denotes the number of agents in the state $\alpha$ and $N$ the number of agents in the system. As usually used in opinion dynamics models \cite{nyczka2012,nowak2021}, $\sum C_{\alpha} = 1$ and we distinguish the following phases:

(i) The disordered phase, when all agent states are of a similar concentration in the system.

(ii) The ordered phase, when one agent state is majority over the others.

(iii) The consensus phase is when the system reaches a particular ordered phase where all agents have the same state.

It is essential to note what we define order from the opinion dynamics perspective to describe order-disorder transitions. Thus, by order, we refer to a macroscopic pattern in which we could find a majority opinion state. We do not refer to the common idea related to public order as the absence of criminal or political violence in society. Moreover, these definitions are convenient because they allow us to identify a macroscopic state with the agents' state and the system's dynamics.\\

{\it (b) Energy}. Now, we introduce a global quantity that allows us to analyze and interpret the system based on the macroscopic states that emerge from the agents' states of the system. Hence, we introduce a Potts-like energy function \cite{wu1982},

\begin{equation}
        E[\alpha] = - \dfrac{1}{2vN}\sum_{i=1}^N\sum_{j\in V_i} J_{ij}\delta(\alpha_i,\alpha_j),\label{eq3}
\end{equation}

\noindent where the symbol $\sum_{j\in V_i}$ means the sum over all neighbors $j$ of the agent $i$ with the same state $\alpha$. Here $J_{ij}$ is called the interaction strength. However, we will take $J_{ij}=1$ (for all $i$ and $j$), due to the characteristics original model. In other words, in this work, we do not consider different interaction strength values. $\delta\left( \alpha_i,\alpha_j\right) $ is a Kronecker delta, i.e., $\delta\left( \alpha_i,\alpha_j\right)=1$ if $ \alpha_i = \alpha_j $ and zero for all $ \alpha_i \neq \alpha_j $.

Note that the energy functions introduced in the Schelling model \cite{Pollicott2001,goles2011} are like the Ising model because the models have two possible states. For the case of the Sakoda model \cite{medina2017} and this model, it is more natural to use the Potts energy because these models have more than two states.

Besides, we can observe that the energy definition shows the absolute minimum or ground state when all the agents in the system are in the same state. On the other hand, the energy may reach the maximum energy when the system takes a chessboard aspect. This behavior is convenient because it allows us to establish analogies or interpretations of the dynamics.

\begin{figure}
    \includegraphics[width=0.48\textwidth]{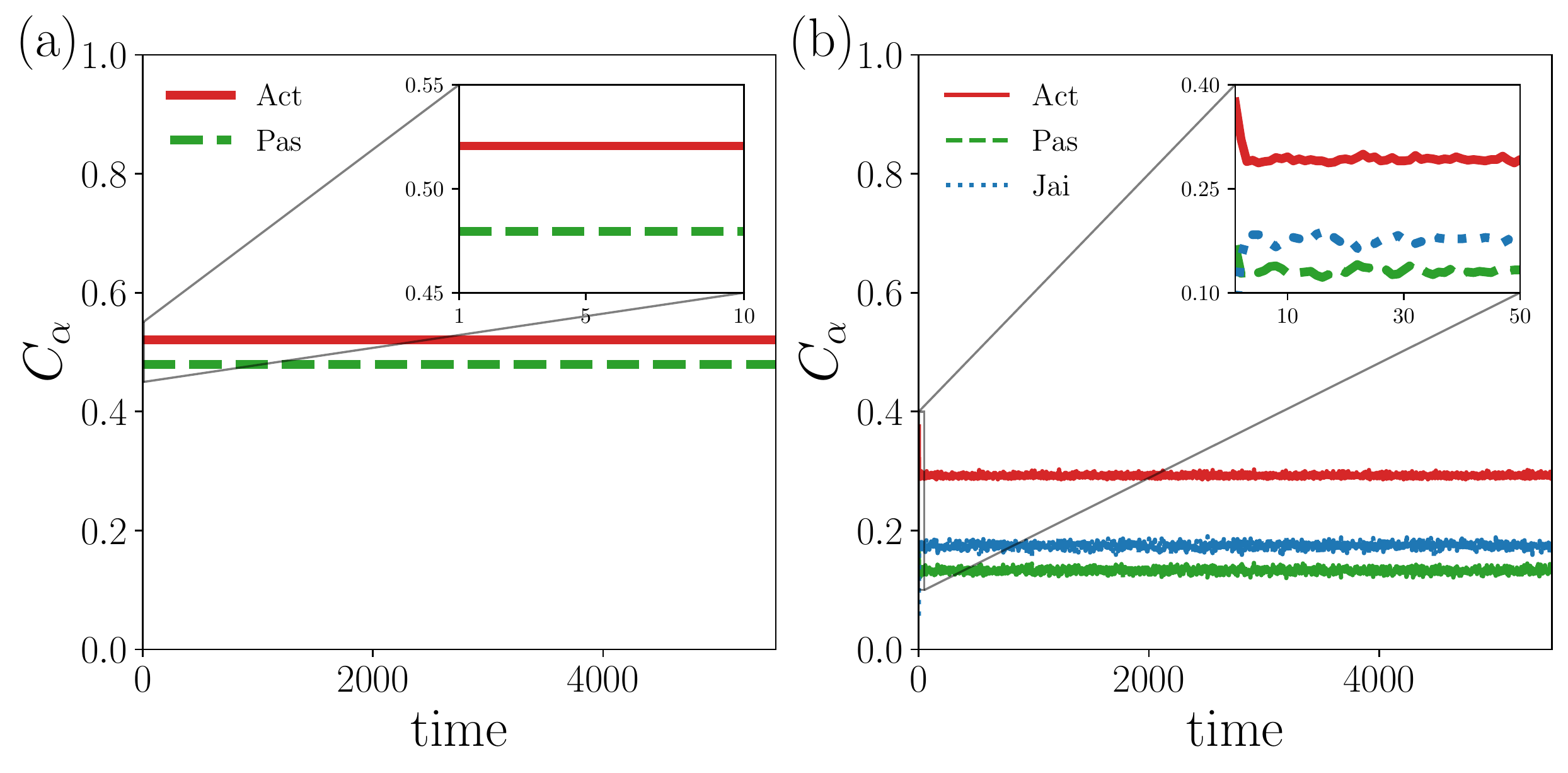}
    \caption{\label{fig_14}Agents' concentration variation on time. Figure (a) corresponds to a system without police officers and Fig. (b) a system with police officers. In both cases, the system reaches a stationary state quickly. }
\end{figure}

\begin{figure*}
    \includegraphics[width=0.95\textwidth]{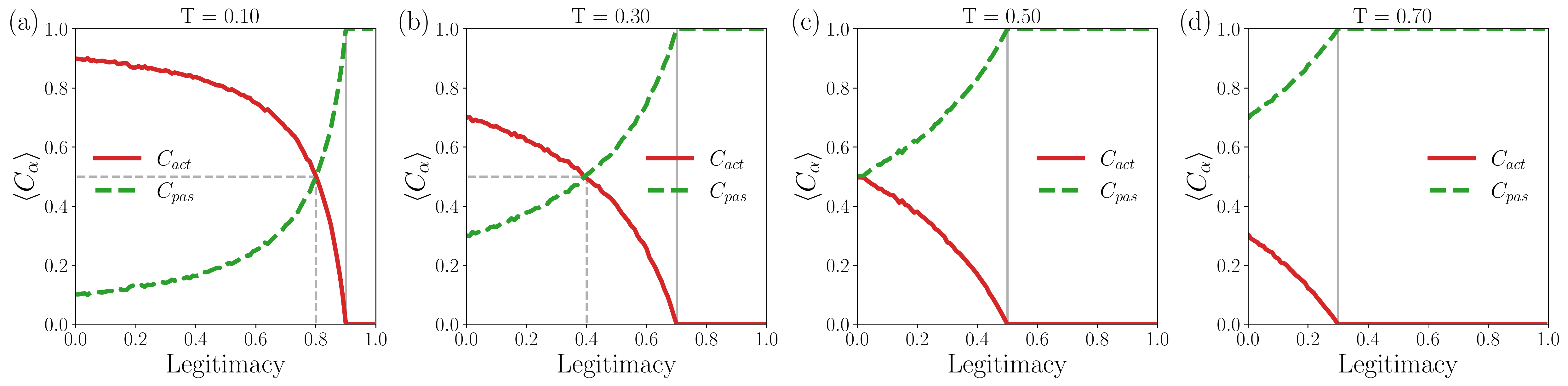}
    \caption{\label{fig_2}Concentration variations for different threshold fixed values in a system without police officers. With low values to legitimacy, the active agents are predominant, but as legitimacy increases, the passive agents dominate. There are points when active and passive agents have a similar concentration (segmented lines) and when all the agents in the system are passives (solid lines). As the threshold increases, we can see a translation of these points suggesting a transition. This figure shows simulations results for a one-dimensional lattice with $N=2^{10}$ sites, Moore neighborhood, and vision one. However, the results obtained for all visions, neighborhoods, and sites considered collapse on the same curve.}
\end{figure*}

\section{Simulations and phase diagrams}\label{diagrams}

This paper aims to characterize the Epstein's model for a civil disorder in a one-dimensional lattice with periodic boundary conditions as a first approximation to understand the dynamics of social protests. Thus, we perform simulations on a one-dimensional lattice with $N = 2^8$ and $N=2^{10}$ sites, considering a system without and with police officers. The first one has agents with two possible states, active and passive. The second one has police officers then the agents can be active, passive, or jailed.

Furthermore, to study the effects of interactions in the system dynamics, we consider agents with visions one and seven interacting in Moore and random neighborhoods. Note that vision determines the neighborhood's size, so when we consider a one-dimensional lattice, the vision represents the number of pairs of agents to consider to evaluate an agent's state switch. For example, when the agent's vision is one, the Moore neighborhood of the agents consists of its nearest neighbors. Thus each agent considers two sites, one to their left and one to their right. When the agent's vision is seven, the Moore neighborhood of the agents counts seven sites on the left and seven sites on the right, with fourteen agents in total.

On the other hand, a random neighborhood is when an agent can randomly select other agents to form their neighborhood. Then, when the agent's vision is one, it chooses two agents randomly. When its vision is seven, the agent can randomly select fourteen agents from the lattice. 

We have not used Epstein's motion rule from the original model in this work. So the agents occupying the whole of the one-dimensional lattice and the system can enter a stationary regime, as shown in Fig. \ref{fig_14}. As we can see, the dynamics converge quickly around the same value for the agents' concentration in any state and remains constant on average after some time. For this reason, in this paper, we can study the asymptotic properties of the dynamics and characterize the model on the steady-state. In this way, we use $20$ realizations with $5500$ time steps for all the study cases. In each realization, all agents have different initial positions and state variables. Then, we discard the first $500$ time steps to obtain a steady state. Finally, we calculated the average quantities over $20$ realizations to characterize the model.

\subsection{System without police officers}

When we study the system without police officers, the state switch equation (\ref{eq1}) changes to 
    \begin{equation}
      H(1-L)>T. \label{eq4}
    \end{equation}
    
Hence, the agents' state only depends on its local parameters and is independent of their neighborhood. Therefore, changes in the system's dynamics depend on the threshold and the initial simulation's conditions. To study the whole system, we run simulations for threshold and legitimacy values between $0.00$ and $0.99$ with a step of $0.01$ for both variables.

\begin{figure}
    \includegraphics[width=0.45\textwidth]{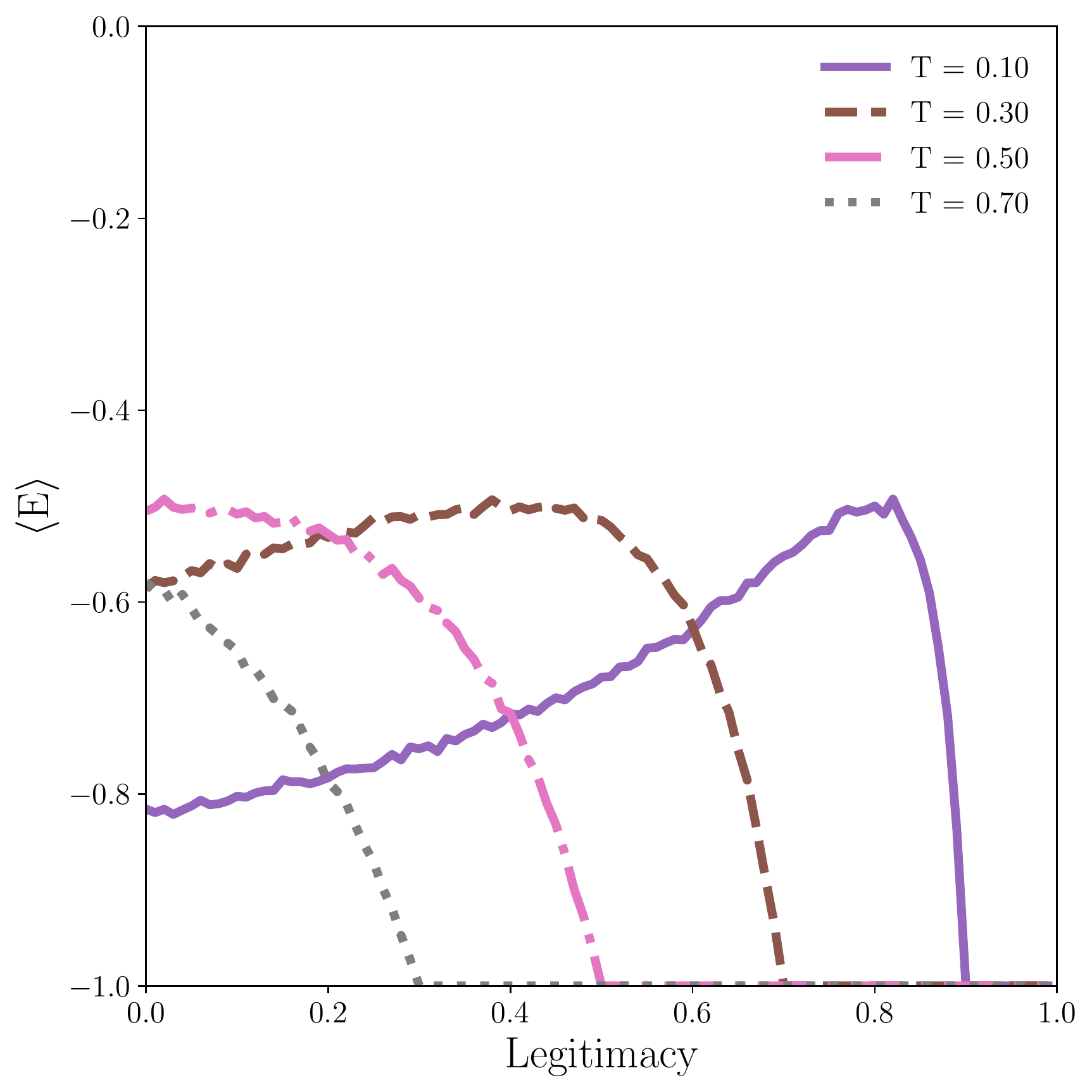}
    \caption{\label{fig_3}The global average energy variations for different threshold fixed values in a system without police officers. The energy shows two minimum values. The first is a local minimum, where the active agents predominate in the system. The other one is a global minimum or the ground state, and all agents in the system are passive. On the other hand, the system reaches a maximum value when the active and passive agent concentrations are similar. This figure shows simulations results for a one-dimensional lattice with $N=2^{10}$ sites, Moore neighborhood, and vision one. However, the results obtained for all visions, neighborhoods, and sites considered collapse on the same curve.}
\end{figure}

To obtain a first idea of the model dynamics, we study the concentration and energy variations for different threshold fixed values. We observe the variation of the concentration of agents when the legitimacy increases in Fig. \ref{fig_2}. With low values to legitimacy, the active agents are predominant. Then as legitimacy increases, the passive agents are predominant. When the threshold is $T=0.10$, the system's dominant state changes, as shown in Fig. \ref{fig_2}(a). Note that when the legitimacy is $L< 0.80$, the active agents predominate, when $L=0.80$, the concentration for two states are similar, and when $L>0.80$, the passive agents are dominant. For $L \geq 0.90$, all agents of the system are in the passive state. We can see a translation of the point of concentration similarity and the point when all agents of the system are passive states when the threshold increases in \ref{fig_2}(a), \ref{fig_2}(b), \ref{fig_2}(c), and \ref{fig_2}(d) figures. The translations of these points indicate transitions in the system. It is important to note that all showed results collapsed on the same curves for all visions, neighborhoods, and sites considered in our simulations. This behavior is because the state switch equation (\ref{eq4}) is independent of the neighborhood and indicates the system's dynamic depending on the threshold and the initial simulation's conditions.

We present the global average energy versus legitimacy with different threshold values in Fig. \ref{fig_3}. When the threshold is $T=0.10$, the energy starts with a lower value of around $\left<E\right>\approx -0.8$. As the legitimacy increases, the energy reaches a maximum value around $\left<E\right>\approx -0.5$ when the legitimacy is $L=0.80$. Next, the energy converges to the minimum value when the legitimacy is $L \geq 0.90$. The initial energy value is a local minimum and indicates when the active agents predominate. Then, when the agents' concentrations are similar, the energy reaches a maximum. Finally, the energy minimizes when the system has only passive agents and reaches the absolute minimum or the ground state.

On the other hand, when the system possesses higher thresholds values, the energy reaches a maximum and then converges to the minimum quickly. Besides, we can see a translation of these energy points as the threshold increases suggesting a transition.

\begin{figure}
    \includegraphics[width=0.45\textwidth]{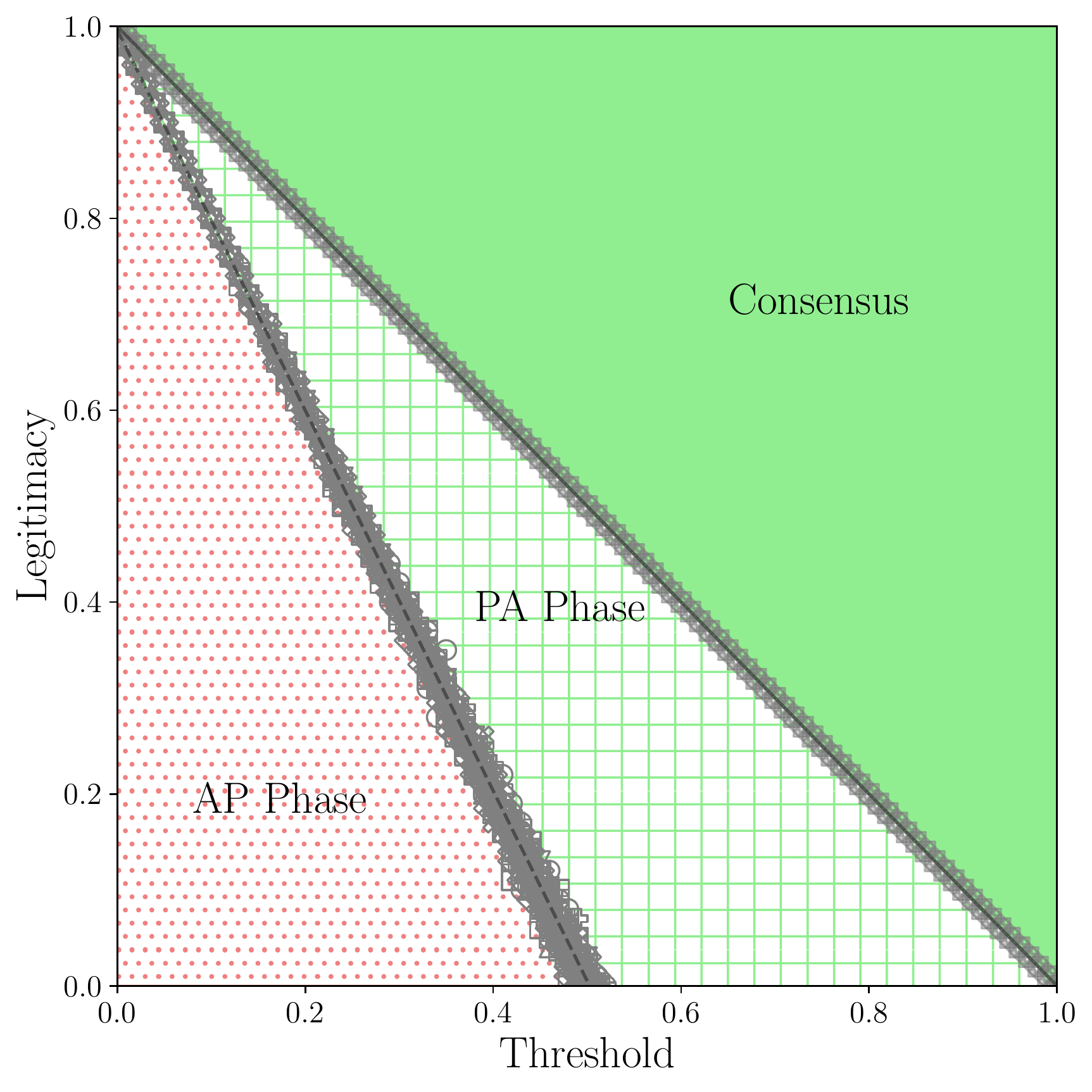}
    \caption{\label{fig_4} Phase diagram for a system without police officers. Phases AP and PA are ordered phases with a majority agent state. Active agents are predominant in the AP phase and passives in phase PA. The dashed line between these two phases shows when the system has similar concentrations, therefore disordered. The consensus phase is a particular ordered phase when all agents are passives. The solid line shows the transition from the order with a majority to consensus. Every point in this diagram corresponds to an $L$ and $T$ value when the concentrations of active and passive agents are similar or when the system reaches a consensus. The points for all visions, neighborhoods, and sites considered collapse on the same curve.}
\end{figure}

To build the phase diagram, we search the coordinates $(T, L)$ where the agents' concentrations are similar, and all agents are passive. As shown in Fig. \ref{fig_4}, these points define the phase boundaries. Following the phases described in opinion models, the results show order-disorder transitions. Phase AP and PA are ordered phases with a majority agent state. Active agents are dominant in phase AP and passives in phase PA. As a result of crossing the dashed line between these two phases, we observe a disordered phase with similarly active and passive agents concentrations. Note that our numerical results are consistent with the assumption for an average agent. We can find the critical legitimacy to obtain equal concentrations of active and passive. As a result, we obtained this dashed line of critical legitimacy $L_c =1 - 2T$. The consensus phase is a particular ordered phase when all agents are passive. The solid line shows the transition from majority order to consensus order.

\begin{figure}
    \includegraphics[width=0.45\textwidth]{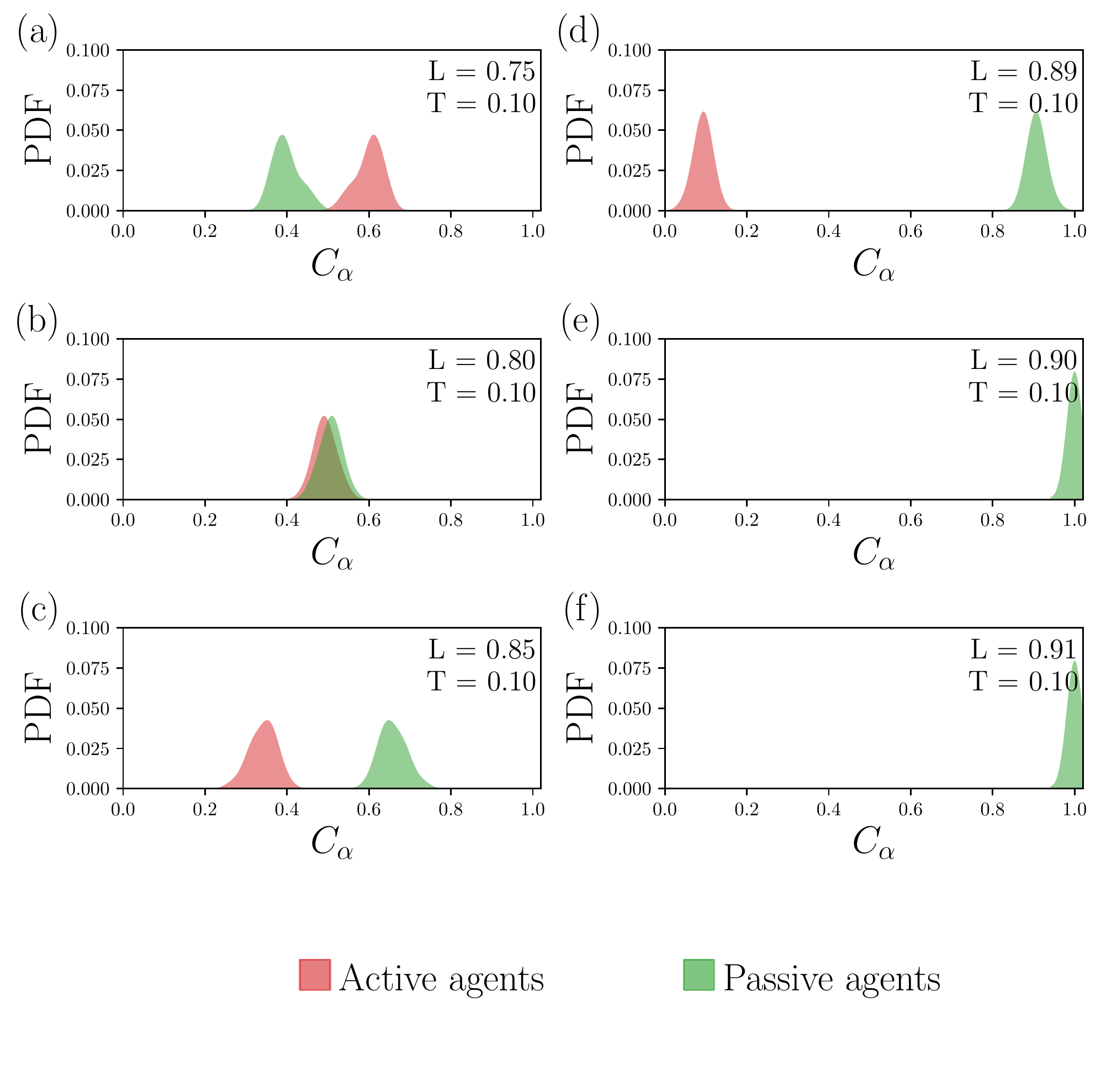}
    \caption{\label{fig_5}Stationary probability density function of the agents' concentration for a system without police officers. We see two transitions when the system increases legitimacy and threshold fixed. A continuous transition from AP phase to PA phase across a disordered phase in $L=0.80$, as seen in figures (a),(b), and (c). Then, a continuous transition from PA phase to consensus, as seen in figures (d), (e), and (f). This figure shows simulations results for a one-dimensional lattice with $N=2^{10}$ sites, Moore neighborhood, and vision one. However, these behaviors are the same for all visions, neighborhoods, and sites considered in our simulations.}
\end{figure}
\begin{figure*}
    \includegraphics[width=0.83\textwidth]{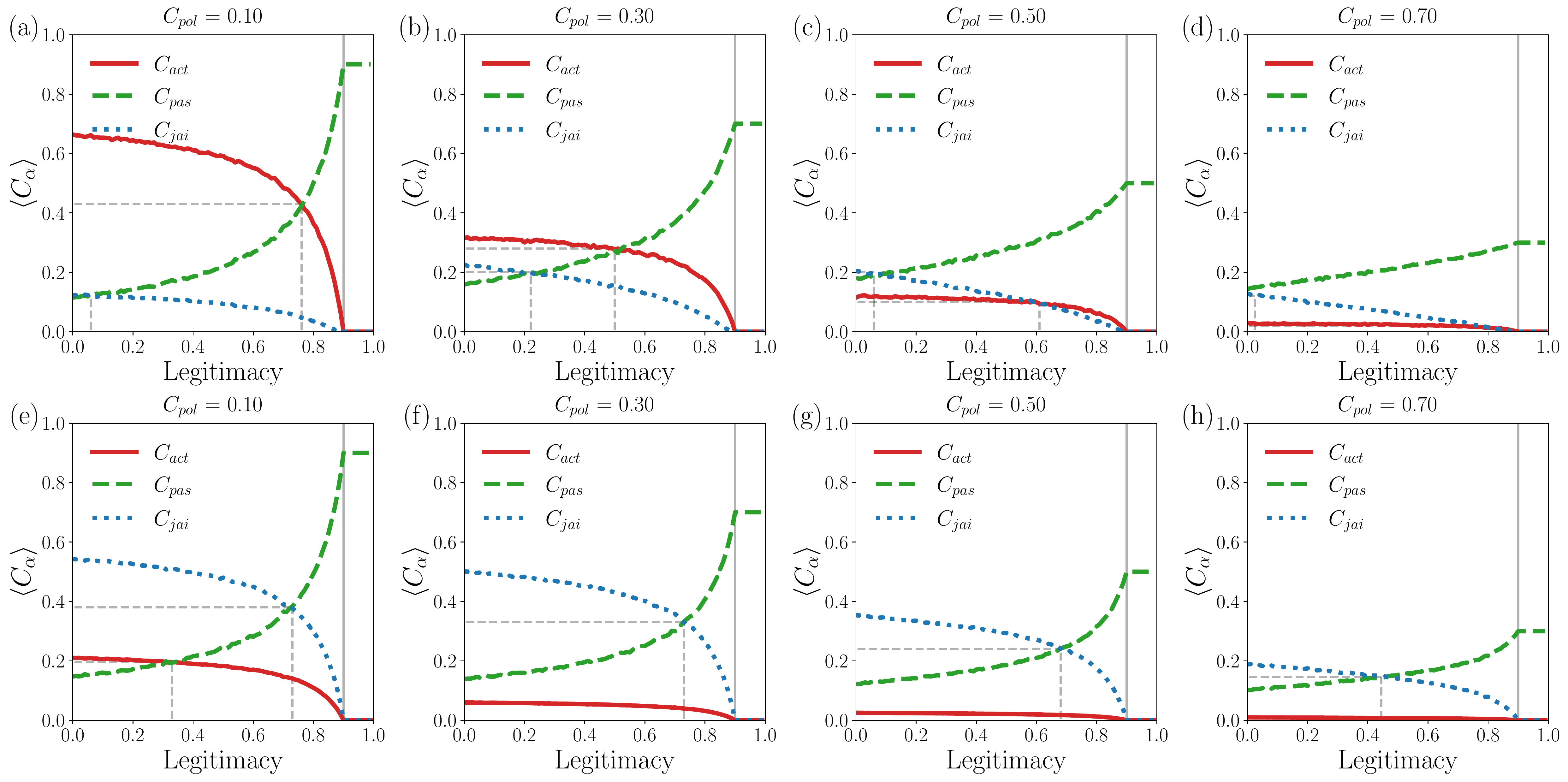}
    \caption{\label{fig_6}Agents concentration variations in a system with police officers for different values of police officers concentrations, legitimacy, vision one, and threshold fixed. When the interactions occur in a Moore neighborhood, the predominance of active agents decreases when the police officers' concentrations increase because of increased jailed agents. As a result, we can see a translation of the concentration similarity points (segmented lines) in figures (a), (b), (c), and (d). These translations show changes of the predominant state and the existence of a point in which the three states of the system are similar, suggesting a phase change. In a random neighborhood, we can see the same dynamics of translation of the concentration similarity point in figures (e), (f), (g), and (h). However, police officers can capture more active agents because of the random selection of their neighborhoods. Thus, the predominance of jailed agents results until the system reaches a high legitimacy. The vertical solid line depends on the fixed threshold and indicates when the system reaches a consensus. This figure shows simulations results for a one-dimensional lattice with $N=2^{10}$ sites, but with $N=2^8$ sites, we observe the same result.}
\end{figure*}

To observe the system transition, we study the stationary probability density function of the agents' concentration. We show distributions for a system with Moore neighborhood with vision one and $T=0.10$ in Fig. \ref{fig_5} because we observe the same behavior independent of the neighborhood, vision and sites considered in our simulations. We can see the system transition from the order with active agents majority in $L=0.75$ to a disordered phase in $L=0.80$, then a change to order with passive agent majority in $L=0.85$, in figures \ref{fig_5}(a), \ref{fig_5}(b), and \ref{fig_5}(c). As for legitimacy increases, we can observe a transition to a consensus in $L=0.90$ in figures \ref{fig_5}(d), \ref{fig_5}(e), and \ref{fig_5}(f).

\begin{figure}
    \includegraphics[width=0.48\textwidth]{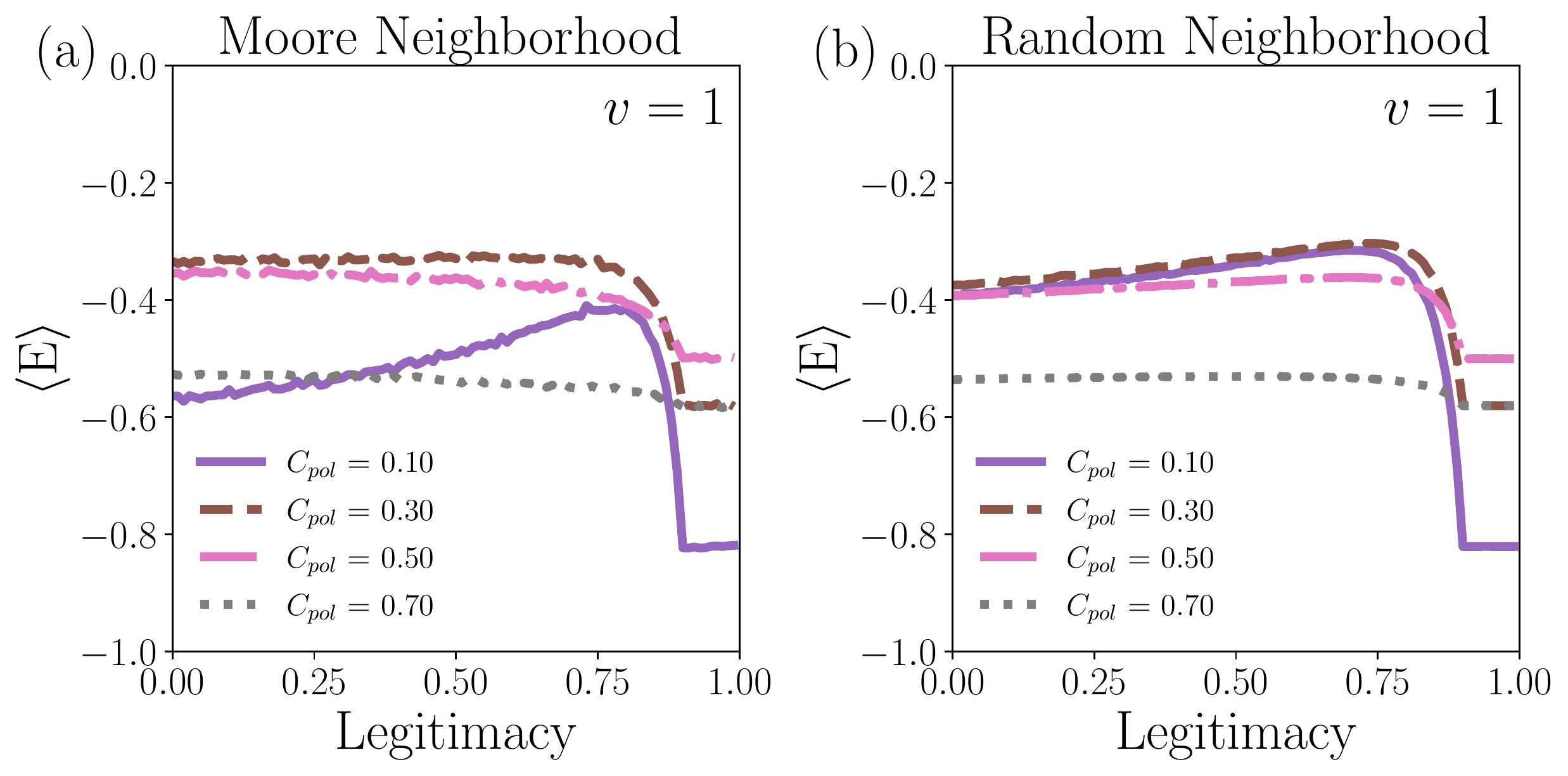}
    \caption{\label{fig_7}The global average energy variations in a system with police officers for different values of police officers concentrations, legitimacy, vision one, and threshold fixed. In a Moore neighborhood, the energy reaches a minimum value for low police officers concentrations and increases as legitimacy increases, as shown in figure (a). When the system reaches a consensus, the energy shows a global minimum. As it increases the police officers' concentration, it is possible to observe that the energy maintains a constant value before reaching a consensus of passive agents. The maximum energy value indicates when the majority agent states have similar concentrations. In a random neighborhood, the energy has a constant value and increases as legitimacy increases, as shown in figure (b). In both cases, the global minimum of energy increase as a police officer's concentration increases. This figure shows simulations results for a one-dimensional lattice with $N=2^{10}$ sites, but with $N=2^8$ sites, we observe the same result.}
\end{figure}

\subsection{System with police officers}

To study the model with police officers, we used legitimacy values and the concentration of police officers from $0.00$ to $0.99$ with a step of $0.01$ for both variables. We vary the concentrations of police officers as an initial condition because its value determines the system's dynamics. The police officers' role is to dissuade a social protest, preventing citizen agents from becoming active agents and arrest active agents in the system producing jailed agents. Besides, the police officers' action depends on the vision, so we study the system separately with two different visions. On the other hand, the police officers' inclusion makes the change of the agents state depends on the state parameters and the neighborhood conditions, as we can see in Eq. (\ref{eq1}). We used a fixed threshold value in $T=0.10$ for these simulations because its role is to determine a limit value to the state switch equation. Furthermore, this value coincides with Epstein's reported value in the five study cases from the original model for a generalized rebellion. Thus, this value allows us to characterize this context with a more significant parameters set.

\begin{figure*}
    \includegraphics[width=0.80\textwidth]{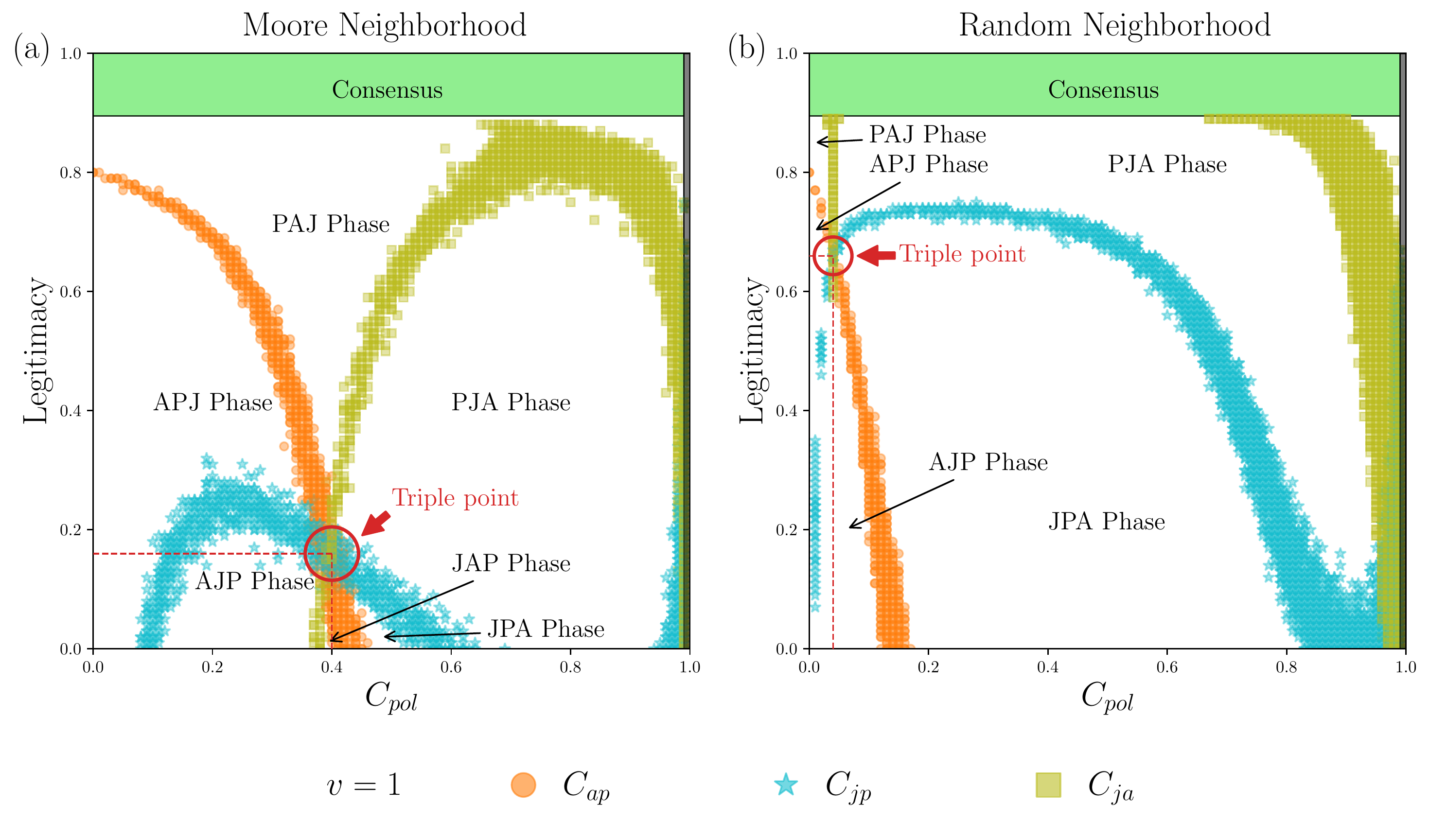}
    \caption{\label{fig_8}Phase diagrams for a system with police officers and vision one. Every point in this diagram corresponds to police officers' concentration and legitimacy value when the different states are at similar concentrations. For example, the $C_{ap}$ points (\protect\marksymbol{*}{orange}) are the points at which the concentrations of active and passive agents are similar. These points determine different ordered phases with a majority, and each label indicates the order of the predominant state. For example, in the PAJ phase, passive agents are predominate, followed by active and jailed agents, and so on. The triple point in these curves' intersection indicates when the three states are in similar concentrations, so we observe a disorder in the system. The system reaches a consensus when legitimacy is greater than or equal to $0.90$. The black region indicates when there are only police officers in the system. With Moore neighborhood show six differents order phases with a majority, one point where the system reaches a disorder, and a consensus phase. With random neighborhoods, order phases with majority change because police officers can capture more active agents and predominantly jailed agents. However, the system reaches a consensus phase in the same conditions because the fixed threshold determines this change. This figure shows simulations results for a one-dimensional lattice with $N=2^{10}$ sites, but with $N=2^8$ sites, we observe the same result.}
\end{figure*}

\begin{figure}
    \includegraphics[width=0.45\textwidth]{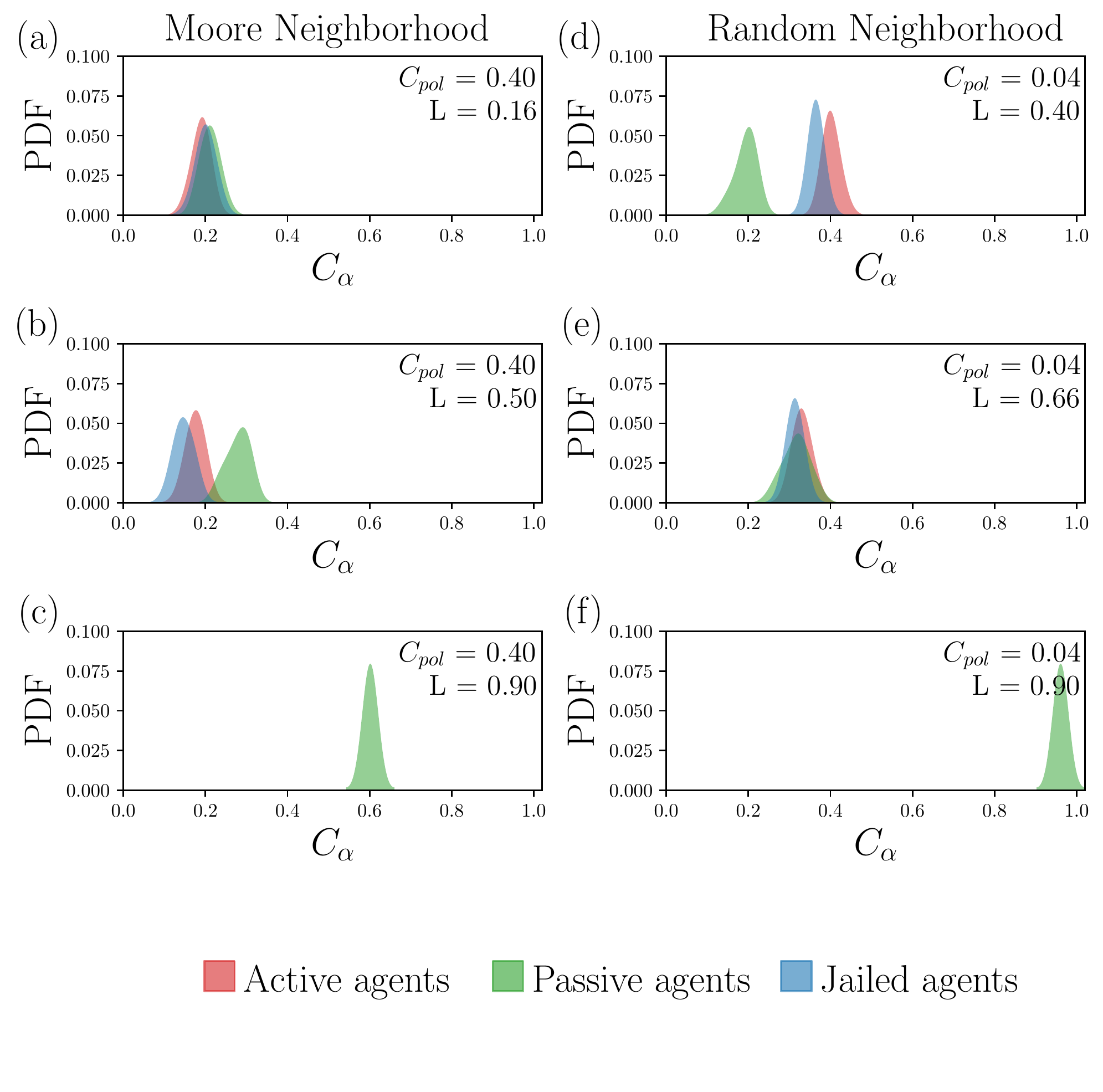}
    \caption{\label{fig_9}Stationary probability density function of the concentration of agents for a system with police officers and vision one. We see two transitions when the system increases legitimacy and fixes police officers' concentration. In the Moore neighborhood, we can see a disordered phase, then order with passive agents majority, and finally, a consensus phase in figures (a), (b), and (c). The same dynamics occur in the random neighborhood, with an order with active agents majority, a disordered phase, and then a consensus phase in figures (d), (e), and (f). This figure shows simulations results for a one-dimensional lattice with $N=2^{10}$ sites, but with $N=2^8$ sites, we observe the same result.}
\end{figure}

\subsubsection*{Results with vision one}

We show the agents' concentration variations for different values of police officers' concentrations and vision one in Fig. \ref{fig_6}. When the interactions occur in a Moore neighborhood, we observe a variation of the concentration of agents when the legitimacy increases in figures \ref{fig_6}(a), \ref{fig_6}(b), \ref{fig_6}(c), and \ref{fig_6}(d). With low values to legitimacy, the active agents are predominant. Then as legitimacy increases, the passive agents are dominant. The jailed agents' concentration depends on the police officers' concentration. Then their variations only occur as police officers' concentration increases and produce a change of active agents concentration. With interactions in the random neighborhood, the jailed agent concentrations have predominant values for low legitimacy, as we can see in figures \ref{fig_6}(e), \ref{fig_6}(f), \ref{fig_6}(g), and \ref{fig_6}(h). Then, passive agents are predominant as legitimacy increases. The active agent concentrations depend on the police officers' concentrations and decrease as the number of police officers in the system increases.

Now, we can observe the global average energy versus legitimacy for a Moore and random neighborhood in figures \ref{fig_7}(a) and \ref{fig_7}(b), respectively. On the one side, when the interactions occur in the Moore neighborhood, and police officers' concentration equals $C_{pol}=0.10$, the energy starts around $\left<E\right>\approx -0.6$. Next, it has an increasing behavior to around $\left<E\right>\approx -0.4$, and a legitimacy value is close to $L=0.80$. Then, the energy decreases quickly to the lower value $\left<E\right>\approx -0.8$. We observe the local energy minimum when active agents are predominant. Then, the energy maximum shows when the agents' states have similar concentrations. In particular, the active and passive agents concentration is approximately $0.4$, and jailed agents and police officers concentrations are close to $0.10$. Finally, we see the global energy minimum when all agents are passive. Note that this global minimum is not absolute because the police officers' concentration equals $C_{pol}=0.10$. As the police officers' concentration increases, the energy maintains a constant value before reaching the minimum energy value. Note that this global energy minimum increases as a police officer's concentration. On the other side, for a random neighborhood, the energy started around $\left<E\right>\approx-0.4$, with a police officer concentration equal to $0.10$, $0.30$, and $0.50$. Next, the energy had an increasing behavior until it reached a maximum around $\left<E\right>\approx-0.3$ and eventually converged rapidly to different energy minima. We observe similar behavior in both neighborhoods for police officer concentration is $C_{pol}=0.70$. Furthermore, the global minimum of energy increases as a police officers' concentration increases for all cases.

As we noticed in the results for a system without police officers, in the systems with police officers, there are also points where the state concentrations are similar. Their positions move as the police officers' concentrations increase. These translations suggest a change in the state predominant in the system. Furthermore, there is a point when the three states of the systems are similar, configuring the order-disorder transitions like opinions models \cite{nyczka2012,nowak2021}. 

To verify this idea, we search for phase boundaries defined by the points in which the agents' concentrations are similar and when all agents reach a passive state. Every point corresponds to police officers' concentration and legitimacy value and depends on a pair of similar agent states. Thus, the $C_{ap}$ coordinate (\protect\marksymbol{*}{orange}) is when the concentrations of active and passive agents are similar. The $C_{ja}$ coordinate (\protect\mystar) indicates similarity in the jailed and passive state, and the $C_{ja}$ point (\protect\marksymbol{square*}{olive!50!}) when jailed and actives agents are similar concentration. Each point formed a curve defining different regions on the phase diagram shown in Fig. \ref{fig_8}. For both the Moore neighborhood in Fig. \ref{fig_8}(a) and the random neighborhood in Fig. \ref{fig_8}(b), we observe phases classified according to the transitions described for the system without police officers. There are six ordered phases with a majority state. Each one has a label indicating the order of the predominant state. For example, the PAJ phase has dominant passives agents, followed by active and the jailed agents, and so on for the other phases. The system reaches a consensus in the passive phase when all agents are passives and legitimacy equal $0.90$. This value is determined by the threshold value selected. The black region indicates when there are only police officers in the system. 

The system reaches a disordered phase labeled a triple point when the active, passive, and jailed states are in similar concentrations. The position of this point and the regions of the phases depends on the neighborhood. Most phases have an observable area for the Moore neighborhood. The part where jailed agents predominate is minor because police officers can only capture active agents among their nearest neighbors. In contrast, police officers are more likely to catch an active agent in a random neighborhood. Thus, we can observe a translation of the triple point and an increase in the regions' size with predominant jailed agents and a decrease in the areas where active agents are dominant. It is important to note that now it is more difficult to see the order-disorder transition. Unlike the system without police officers, where we observe the transition for all parameters, in the system with police officers, the disordered phase is only a point in this diagram. 

We can observe the system transition with the stationary probability density function of the agents' concentration in Fig. \ref{fig_9}. As for legitimacy increases, the system changes from the disordered phase to an ordered phase with a passive state majority. It then reaches a consensus phase in a Moore neighborhood in figures \ref{fig_9}(a), \ref{fig_9}(b), and \ref{fig_9}(c). For a random neighborhood, the transition from the ordered phase with the active state to a disordered phase, then the consensus phase, as we see in figures \ref{fig_9}(d), \ref{fig_9}(e), and \ref{fig_9}(f). The final passive agents' concentration depends on the police officers's concentration fixed to observe the transition. So, in the Moore neighborhood, the final concentration is $0.6$ and, in a random neighborhood, the passive agents' concentration is $0.9$.

\begin{figure*}
    \includegraphics[width=0.82\textwidth]{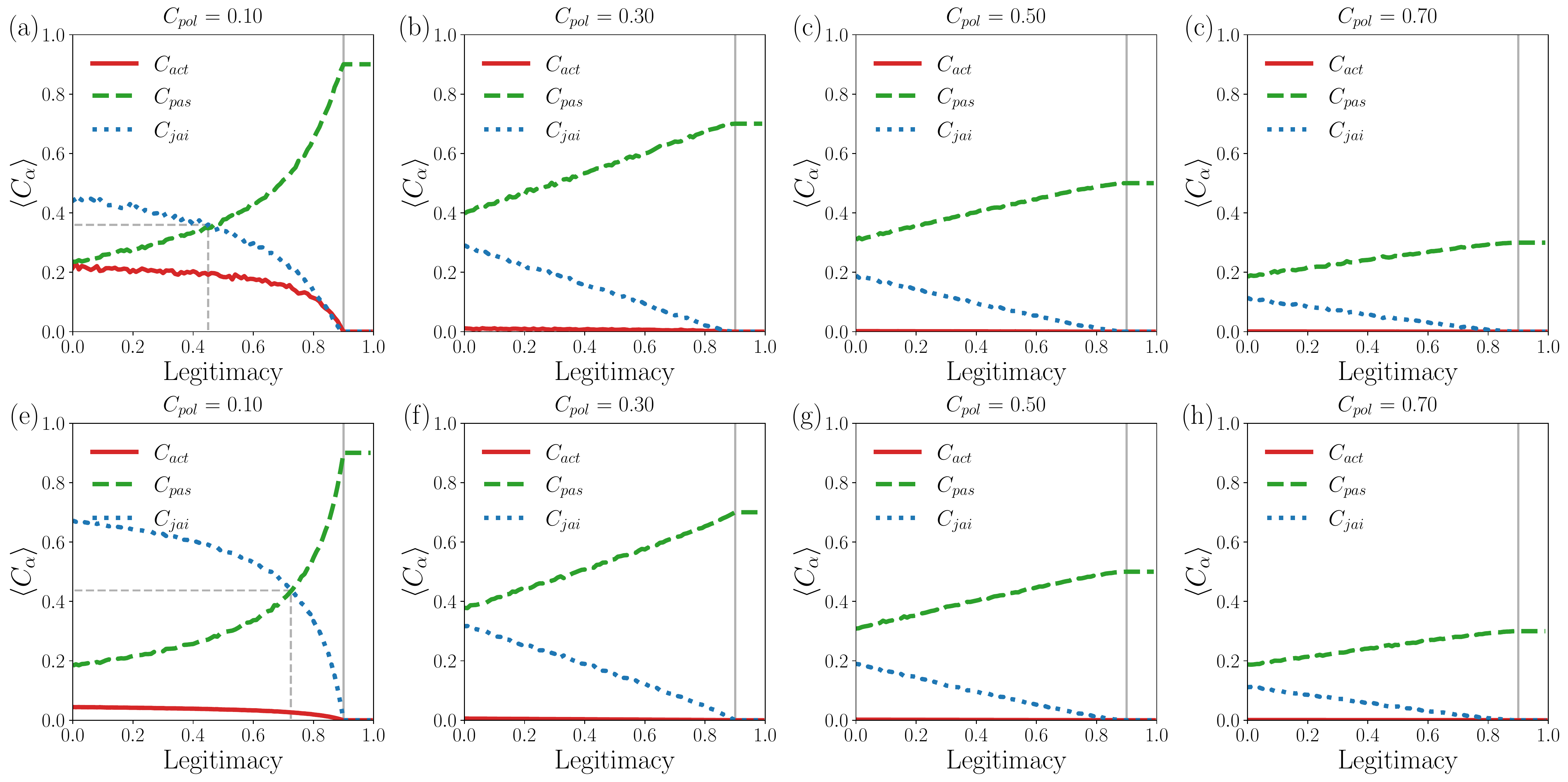}
    \caption{\label{fig_10}Agents concentration variations in a system with police officers for different values of police officers concentrations, legitimacy, vision seven, and threshold fixed. The first and second-row figures show interactions in a Moore and random neighborhood respectably. We can see the predominance of jailed agents with low police officers concentrations in Moore and random neighborhoods in figures (a) and (e). Note that the concentration of jailed agents is more significant in the random neighborhood because police officers can capture more agents. However, as the police officers' concentrations increase, we can see the predominance of passive agents for both kinds of neighborhoods. Besides, we observe the same dynamics of translation of the concentration similarity (segmented lines) as in a system with vision one but only for low police officers concentrations. The vertical solid line depends on the fixed threshold and indicates when the system reaches a consensus. This figure shows simulations results for a one-dimensional lattice with $N=2^{10}$ sites, but with $N=2^8$ sites, we observe the same result.}
\end{figure*}

\begin{figure}
    \includegraphics[width=0.48\textwidth]{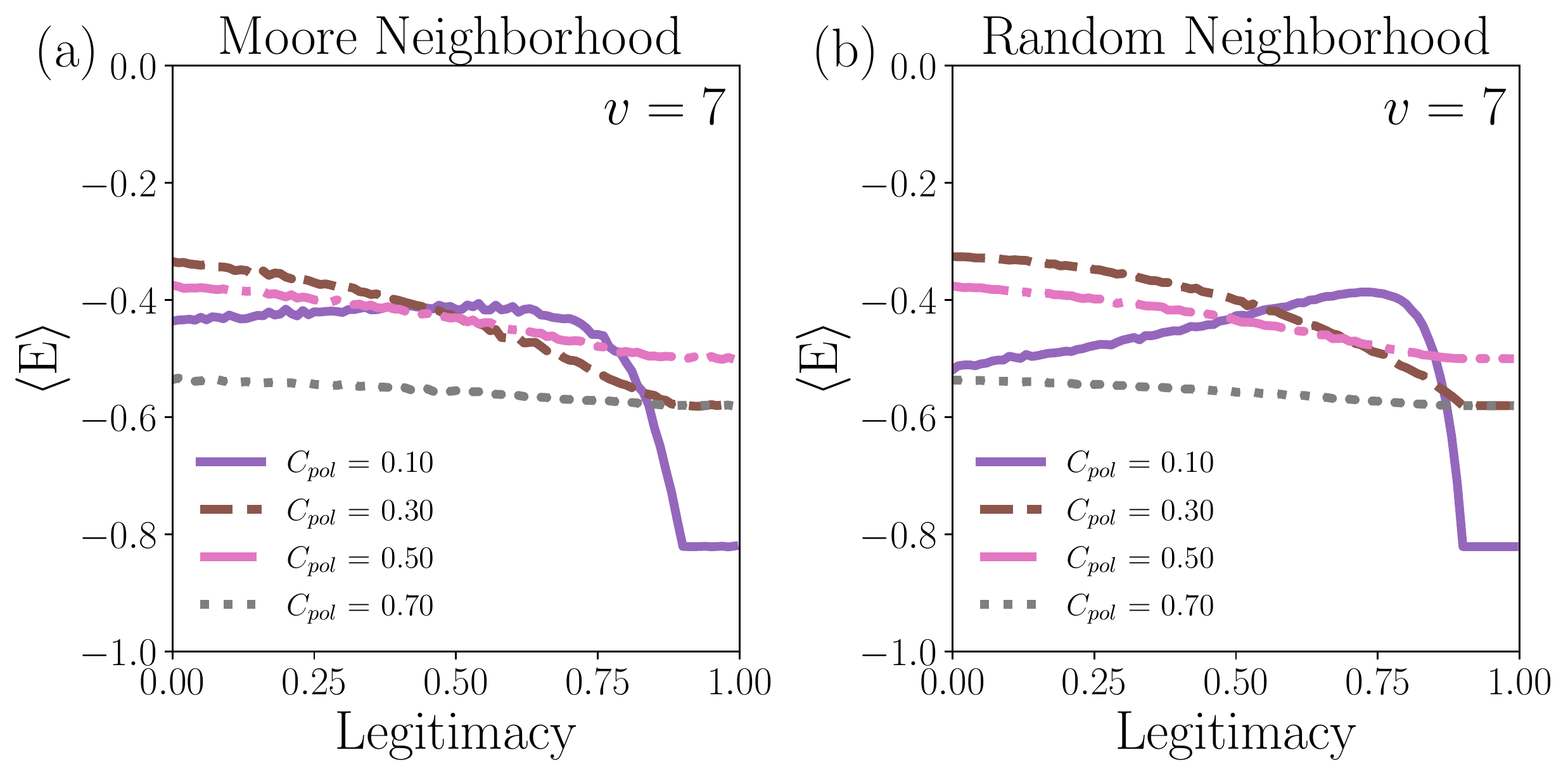}
    \caption{\label{fig_11}The global average energy variations in a system with police officers for different values of police officers concentrations, legitimacy, vision seven, and threshold fixed. As for legitimacy increases, the energy reaches a global minimum in Moore and random neighborhoods. However, the energy only shows a maximum with police officers' concentration values lower or equal to $0.10$. Then, as police officers' concentration increases, the energy constantly decreases until the system reaches a consensus. Besides, in both cases, the global minimum of energy increase as a police officer's concentration increases, as we can see in figures (a) and (b). This figure shows simulations results for a one-dimensional lattice with $N=2^{10}$ sites, but with $N=2^{8}$ sites, we observe the same result.}
\end{figure}

\subsubsection*{Results with vision seven}

We show the agents' concentration variations for different police officers' concentrations and vision seven in Fig. \ref{fig_10}. When the interactions occur in a Moore neighborhood, we can see a variation of the concentration of agents when the legitimacy increases in figures \ref{fig_10}(a), \ref{fig_10}(b), \ref{fig_10}(c), and \ref{fig_10}(d). With low legitimacy, active agents exist, but jailed agents are predominant. Then as legitimacy increases, the passive agents are dominant. As police officers' concentration increases, the active agents' concentration minimizes. Passive agents increase constantly, and jailed agents decrease. 

With interactions in the random neighborhood, the jailed agent concentration has predominant values for low legitimacy, as we can see in figures \ref{fig_10}(e), \ref{fig_10}(f), \ref{fig_10}(g), and \ref{fig_10}(h). Furthermore, the active agent's concentration starts initially with a low value, and passive agents are predominant as legitimacy increases. As police officers' concentration increases, the concentration of active agents disappears, passive agents increase constantly, and jailed agents decrease. This behavior indicates a change in the importance of the police officers' role to dissuade a protest. When the agents have vision seven, the relevant police officers' role is to prevent citizen agents from becoming active and prevent the emergence of a protest.

Now, we can observe the global average energy versus legitimacy for a Moore and random neighborhood in figures \ref{fig_11}(a) and \ref{fig_11}(b), respectively. On the one side, when the interactions occur in the Moore neighborhood, and police officers' concentration equals $C_{pol}=0.10$, the energy starts close to $\left<E\right>\approx -0.4$. Next, it maintains a constant behavior until the energy decreases quickly to the lower value of around $\left<E\right>\approx -0.8$. On the other side, in the random neighborhood, and police officers' concentration equals $C_{pol}=0.10$, the energy has a lower value, around $\left<E\right>\approx -0.3$. Then increases to reach a maximum value close to $\left<E\right>\approx -0.4$ and decreases quickly to the lower value around $\left<E\right>\approx -0.8$. Note that, in both cases, we observe the initial energy value when jailed agents are predominant. Then, the energy maximum shows when the jailed and passive agents concentration is approximately $0.4$, and police officers concentration is $0.10$. Finally, we see the global energy minimum when all agents are passive. However, in both neighborhoods, as police officers' concentrations increase, the energy has a decreasing behavior until reaching the lower energy value. We note that the global energy minimum increases as a police officer's concentration for all cases.

\begin{figure*}
    \includegraphics[width=0.80\textwidth]{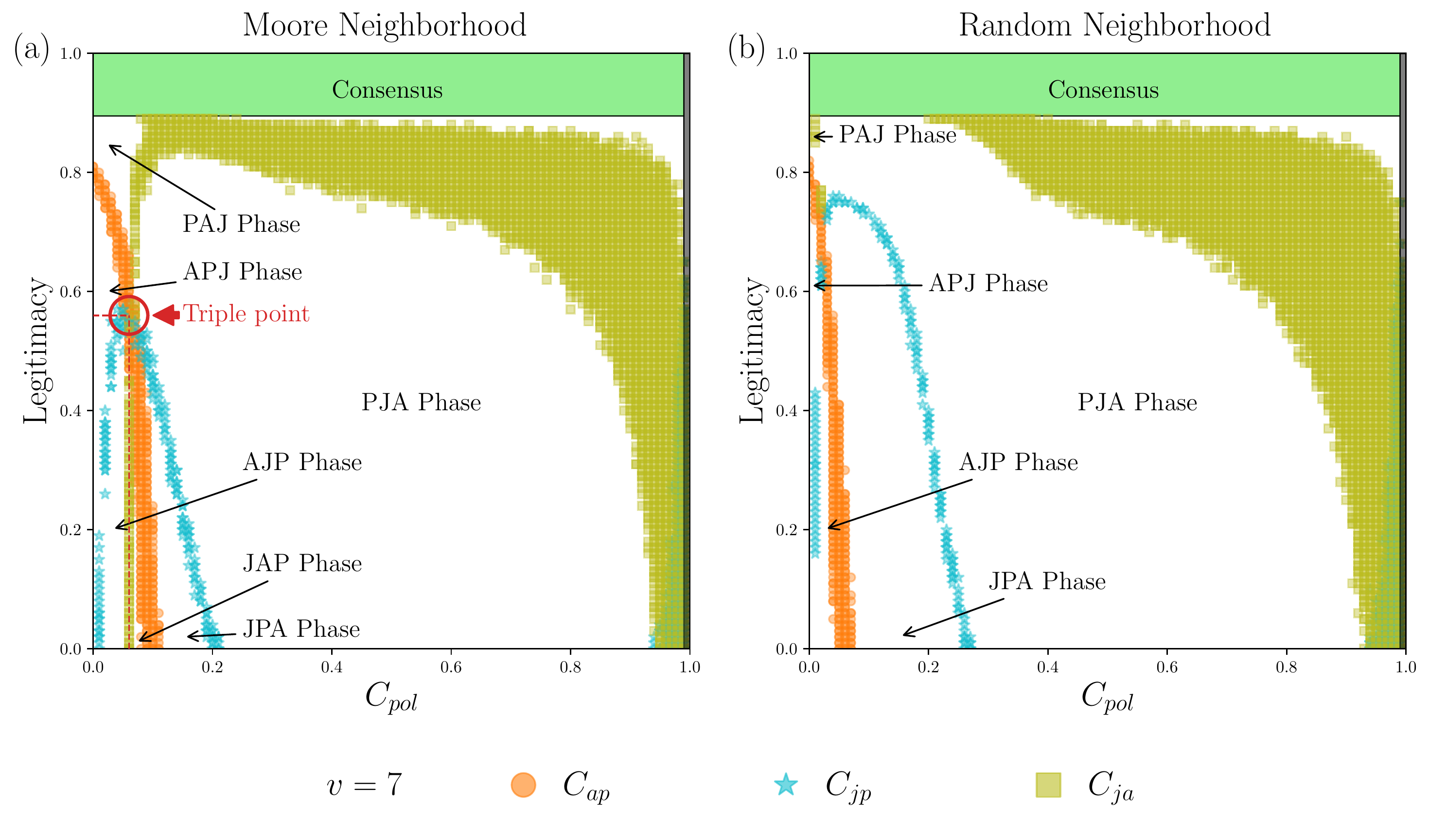}
    \caption{\label{fig_12}Phase diagrams for a system with police officers and vision seven. Every point in this diagram corresponds to police officers' concentration and legitimacy value when the different states are at similar concentrations. We can see the same phases and transitions order-disorder from the system with vision one in the Moore neighborhood. Nevertheless, the size of each phase change. With random neighborhood, the disordered phase and the JAP phase disappear due to increased police officers' action with vision seven, as shown in figure (b). In both cases, the system reaches a consensus when the legitimacy is greater than or equal to $0.90$ because the fixed threshold determines this change. The black region indicates only police officers in the system. This figure shows simulations results for a one-dimensional lattice with $N=2^{10}$ sites, but with $N=2^8$ sites, we observe the same result.}
\end{figure*}

\begin{figure}
    \includegraphics[width=0.45\textwidth]{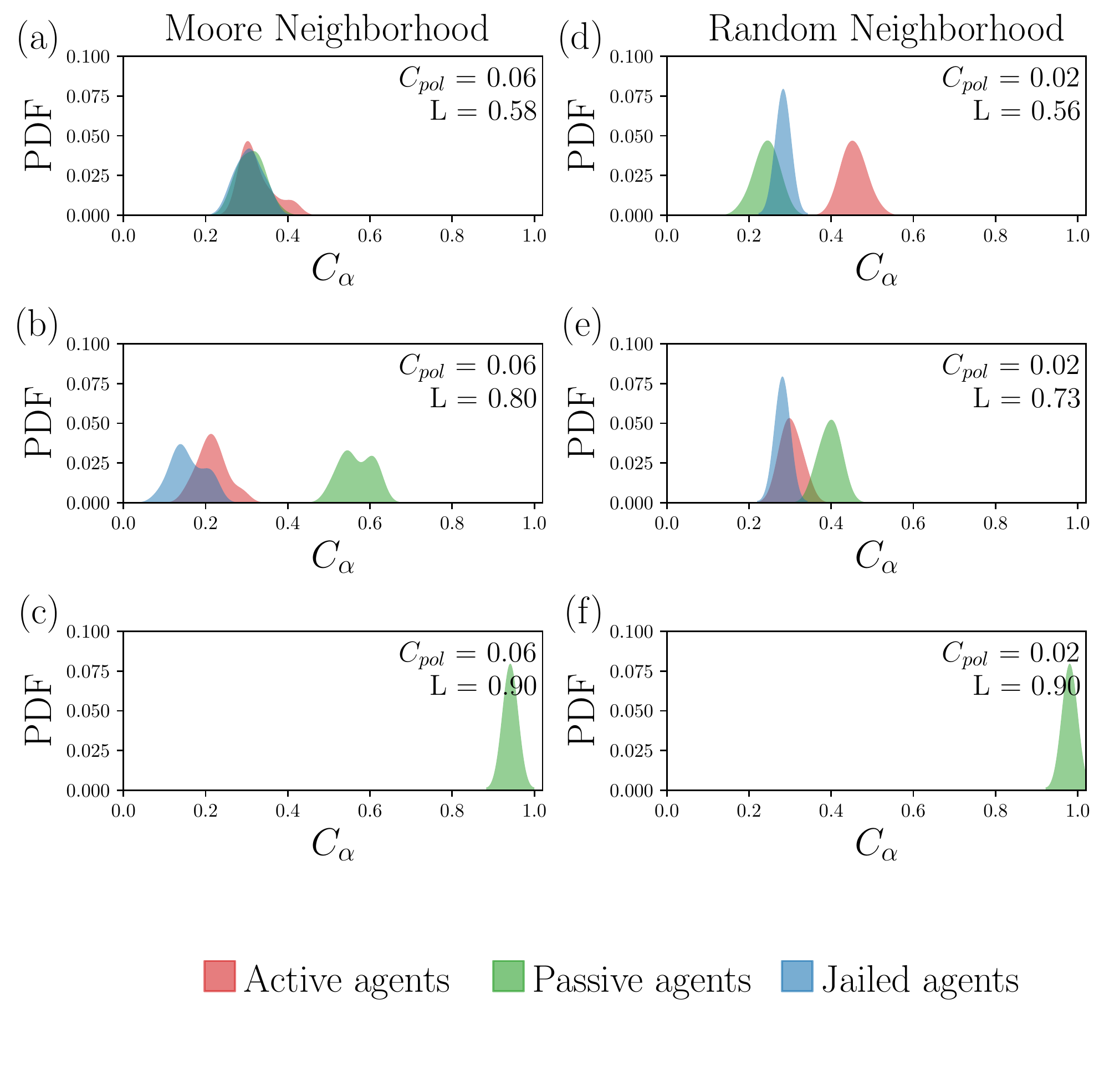}
    \caption{\label{fig_13}Stationary probability density function of the concentration of agents for a system with police officers and vision seven. We can see two transitions in the Moore neighborhood when the system increased legitimacy and fixed police officers' concentration. A disordered phase, then orders with passive agents majority, and finally, a consensus phase in figures (a), (b), and (c). However, in the random neighborhood, only observe order-with majority phases and then a consensus phase in figures (d), (e), and (f). This figure shows simulations results for a one-dimensional lattice with $N=2^{10}$ sites, but with $N=2^8$ sites, we observe the same result.}
\end{figure}

For low police officers concentrations, points where the state concentrations are similar exist, suggesting changes in the predominant state in the system. To search for a point where the system shows an order-disorder transition, we built a phase diagram. Every point corresponds to police officers' concentration and legitimacy value and depends on a pair of similar agent states. Then, we can observe the phases formed for the system with police officers's vision seven in Fig. \ref{fig_12}. We see the same six phases observed in the system for the Moore neighborhood with vision one in Fig. \ref{fig_12}(a). The region's size for every phase changes notably because of the increase in agents' vision and the police officers' activity. As a result, we note that the regions with the predominance of passive agents are more significant than the others. Besides, we can observe a triple point where the disordered phase occurs, the area where the system only has police officers, and a consensus phase with only passive agents. For the random neighborhood, the effect of vision seven is more significant, as shown in Fig. \ref{fig_12}(b). Although the size of the regions dominated by passive agents is similar to those of the Moore neighborhood, we can notice that the JAP phase and the point at which all concentrations are similar disappears. As a result, we observe that there is no order-disorder transition. However, the system change between different order phases with a majority state depending on the police officers' concentration and legitimacy values. It reaches a consensus when all agents are in the passive state at the value of legitimacy is $0.90$ because this depends on the threshold value fixed at the beginning of simulations.

We show the stationary probability density function of the agents' concentration in Fig. \ref{fig_13} to observe the system transition. For the Moore neighborhood, we selected the police officers' concentrations $0.06$ and varied the legitimacy. We can see, as legitimacy increases, the system change from the disordered phase to an ordered phase with a passive state majority in figures \ref{fig_13}(a) and \ref{fig_13}(b). Then reaches a consensus phase in Fig. \ref{fig_13}(c). We selected a lower value of police officers' concentrations for the random neighborhood to observe the possibility of finding similar concentrations for the three states as legitimacy increases. Nevertheless, only find order with majority phases, as shown in figures \ref{fig_13}(d), \ref{fig_13}(e), and the consensus phase in Fig. \ref{fig_13}(f).

\section{Discussion and Concluding Remarks}\label{remarks}

This paper studied the one-dimensional civil disorder model with the whole lattice occupied to characterize their evolution on the steady-state. To do this, we performed extensive numerical simulations of the model with and without police officers, considering visions one and seven in Moore and random neighborhoods to study the effects of interactions on the system's dynamics. We used the agent state concentration and introduced a Potts-like energy function as global quantities to characterize the model.

In the system without police, the dynamics only depend on values assigned as initial conditions. One of them is hardship, a local parameter uniformly distributed between values zero and one for each agent. In the model, this parameter allows for a heterogeneous society of agents. The other two, legitimacy and threshold, are global parameters that we use as control parameters for our simulations. The threshold is a quantity defined as non-negative and determines a limit for an agent's state change.

Given that the product of hardship and legitimacy symbolizes the grievance in the state change equation, we interpret this threshold as a tolerance for grievance. The higher the threshold value, the more disposed agents are to tolerate grievance before rebelling against authority. This parameter takes different values between zero and one, which we interpret as an essential property of a community that depends on its culture or way of life. Thus, we find communities with a low threshold and quickly protest for a grievance and communities with a very high threshold that lives with a minimal grievance and does not rebel against authority. On the other hand, legitimacy is the community's perception of the regime or the system's authority. Thus, a low legitimacy produces more significant grievance, and high legitimacy favors the passivity of the system.

In our results for the variations in the concentration of agents as a function of legitimacy, we observe that active agents are predominant for low values of legitimacy. As legitimacy increases, we find a point where the concentrations of active and passive agents are similar, and then passive agents become predominant. According to the legitimacy variation, the predominance changes depend on the threshold values. We note that passive agents are always dominant for threshold values greater than $0.50$ and coexist with active agents. However, for high legitimacy values, we notice that all agents in the system become passive. These changes in the predominance of a state among the agents are indicators of phase changes. We build the phase diagram based on the concentration points of similar agents or when all agents are passive, and we observe order-disorder transitions. We identify the AP and PA phases as order phases with a majority state. Active agents are predominant in the first and passive agents in the second. We find a disordered phase in which the agents' concentrations are similar when crossing between these phases. When all the agents are passive, the system reaches the consensus phase. We study the transitions with the stationary probability density function of the concentration of the agents, and we observe a typical scenario of continuous transitions \cite{nyczka2012}.

The energy shows one maximum and two minimum values for low threshold values. One of them is a local minimum, where the active agents predominate in the system. The other is a global minimum or ground state, where all agents in the system are passive. The system reaches the maximum value when the concentrations of active and passive agents are similar. In contrast, when the system has higher threshold values, the energy is maximum and quickly converges to a minimum. We can see the translation of these energy points as the threshold increases, allowing us to identify the phase changes. Thus, the local minimum indicates the ordered phase with an active agents majority, the maximum the disordered phase, and the global minimum or ground state is the consensus phase. 

The local energy minimum observed for threshold values greater than zero and less than $0.50$ is a metastable point. The active agents' predominance generates energy and stops the system from reaching the ground state. In other words, the system is not entirely stable due to grievance. However, as legitimacy increases, the system can reach a consensus. On the other hand, when the threshold values are greater than $0.50$, the local minimum disappears and becomes an energy maximum. The maximum is an unstable point for all threshold values because the system can fall to the local or global minimum depending on how legitimacy varies. Since the concentrations of active and passive agents are similar, the system increases energy due to grievance generating a scenario of instability comparable to a polarized society. In this context, we understand polarization as a situation of equal opinion searching for a consensus. It is important to note that the maximum energy that our results show is the maximum possible given the initial conditions of random positioning of the agents. However, it is possible to find an absolute maximum for the energy by positioning the agents deliberately to form the checkerboard appearance. 

Finally, the global minimum is a stable point of the system since all the agents are in the same passive state in the consensus phase. When threshold values exceed $0.50$, the system can reach the global minimum for low legitimacy values. Nevertheless, for threshold values less than $0.50$, we find that reaching consensus requires higher legitimacy values. Indeed, the greater the tolerance for grievance, the less legitimacy is required to reach consensus. This result reflects the existence of societies with high thresholds, in which reaching consensus requires low legitimacy values, unlike other societies in which reaching or maintaining consensus requires high legitimacy.

From these results, we can reveal that the principle underlying the dynamics of the model is a {\it Principle of Minimum Grievance}, equivalent to that observed in the model of worker protest in a factory. This principle allows us to interpret that the system seeks minimal grievance or a consensus. Naturally, reaching and maintaining consensus requires high values of legitimacy or high values of tolerance, the latter being the one that determines the value of legitimacy necessary in a heterogeneous society. Thus, the emergence of protests is due to global conditions of legitimacy or threshold that generate a grievance, increasing the system's energy. This grievance generates energy variations that can lead the system to instability or metastability.

In the system with police, the agents can be active, passive, or jailed, and the dynamic depends on the vision and the neighborhood. For this reason, we study the system separately, considering two different views. We fix the threshold value and consider the legitimacy and concentration of police officers in the system as a control parameter because the police officer's role is to deter a protest. Besides, the agents' state switch depends on the number of active agents and police officers in their neighborhood.

The concentration of agents allows us to identify the predominance of states in the system. We find that active agents predominate in the low legitimacy regime only when the system has vision one and Moore neighborhood. For the rest of the cases, the jailed agents always predominate. As in the system without police, as legitimacy increases, the predominant state changes, and passive agents become the majority. In addition, we again note points where concentrations are similar. For legitimacy values greater than or equal to $0.90$, all agents in the system are passive. It is essential to mention that the increased concentration of police officers in the system facilitates the predominance of passive agents. Therefore, in general, the concentration of passive agents increases as legitimacy increases. In particular, this increase is linear in the system with vision seven because the agents do not switch to the active state as police presence increases in a neighborhood. In addition, when the system reached the legitimacy value in which jailed and passive concentrations are similar, the number of jailed agents began to drop rapidly.

\begin{figure}
    \includegraphics[width=0.38\textwidth]{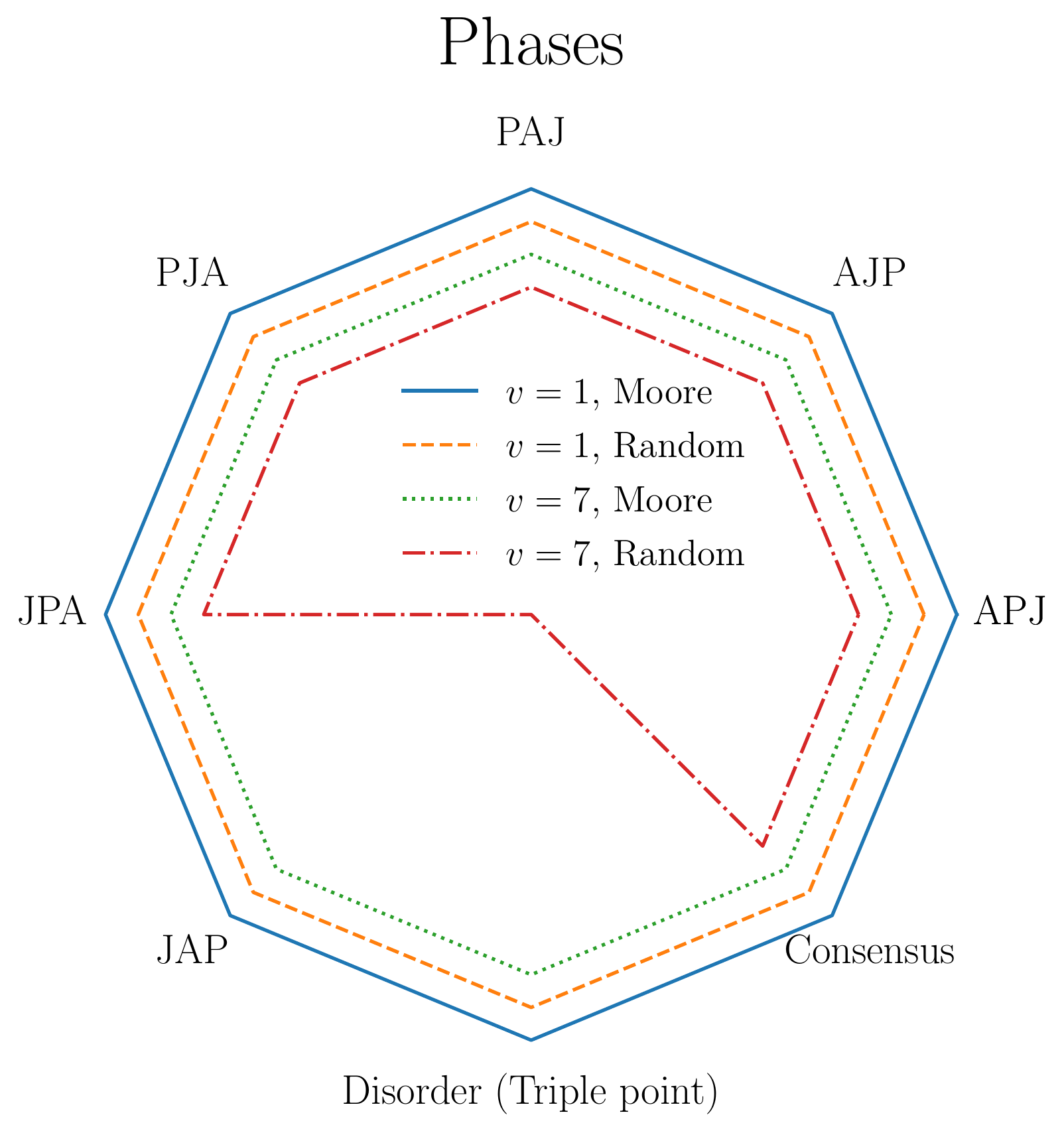}
    \caption{\label{fig_15}Schematic visualization of different phases found at the one-dimensional civil disorder model with police officers on the steady-state. We observe six ordered phases with a majority state, a disordered phase, and a consensus with vision one. With vision seven and Moore neighborhood, we observe the same phases. However, there is neither the disordered phase nor the ordered phase, with most jailed agents followed by active and passive agents (JAP phase) in a random neighborhood.}
\end{figure}

The phase diagrams allow us to identify the domains' limits based on the changes in the predominance of a state in the system. The diagrams show order-disorder transitions for the system with vision equal to one. We identify six ordered phases with a majority state, a disordered phase at a specific point in the diagram, and a consensus phase where all agents in the system are passive. The disorder point's position and the phases' sizes depend on the type of neighborhood. The random neighborhood makes it easier to capture active agents. So, the disorder point requires less concentration of police than in the Moore neighborhood, and the most extensive phases are those with a predominance of passive agents. When the system has vision seven, the differences depend on the neighborhood. In a Moore neighborhood, we observe order-disorder transitions with the same phases observed in the system with vision one. However, in a random neighborhood, we only observe 5 phases of order with a majority, and we do not observe the order-disorder transition. In this neighborhood, police officers significantly increase activity. So, they can capture many active agents in the low legitimacy regime, preventing the emergence of active agents with high legitimacy values. We show a summary of the different scenarios of the system with police officers in steady-state in Fig. \ref{fig_15}. Finally, the study of the stationary probability density functions of the concentration of the officers shows us that the phase changes are continuous at the same as in the system without police.

From the results we obtained for energy, we can note essential differences in behavior according to the concentration of police officers in the system. When the police concentration has values less than or equal to $0.10$, we can see that the energy shows a local minimum, a maximum, and a global minimum. The local minimum shows an ordered state with a majority. We observe the most active agents in the case of vision one and a Moore neighborhood. In the other cases, the jailed agents always predominate. As for legitimacy increases, the energy reaches a maximum when at least two agent states are in similar concentrations. For example, for the system with vision one and a Moore neighborhood, this maximum shows when the concentration of active and passive agents are similar. In the other cases, we observe that the jailed and the passive are in a similar concentration. Therefore, the maximum energy only shows changes between phases of order with a majority and cannot be identified with the state of disorder of the system. Finally, we observe the global energy minimum when all agents in the system are passive, indicating that the system is in the consensus phase. 

When the concentration of police officers has values greater than $0.10$ but less than $0.70$, we observe different behavior of energy depending on the vision and the type of neighborhood. For vision one and a Moore neighborhood, the initial energy is maximum. Then, as legitimacy increases, energy remains constant until the legitimacy value indicates the phase with passive agents majority. Then, the energy decays quickly to the minimum in the consensus phase. The initial energy for vision one and a random neighborhood is a local minimum. We observe a minimal increase until it reaches a maximum and quickly decays to a global minimum. The maximum shows the transition from jailed agents' majority phase to the passive agents' majority phase and then reaches the consensus phase. For vision seven and both neighborhoods, we see that the energy starts at the maximum possible value, indicating a phase where the prisoners predominate. Then, energy steadily decays towards the global minimum in the consensus phase. 

Finally, for regimes where the police concentration is greater than or equal to $0.70$, we see for all cases that the initial energy has a closer value to the minimum. As legitimacy increases, the energy constantly decreases until it reaches the minimum associated with consensus. In general, we can note that the minimum energy for all cases is not an absolute minimum. Despite reaching the consensus phase and all the agents being passive, there are police officers in the system. Therefore, its value depends on the police officers' concentration and their random position on the lattice. This result leads us to conjecture that the energy value will be closer to the absolute minimum for an arbitrary initial configuration with two clusters of police officers and passive agents.

Regarding the observed stability points of energy, we see similarities for the system without police officers. The local energy minimum observed for low police concentration is a metastable point for the same reason as the system without police. For all visions and neighborhoods, this point shows the existence of grievance, either because there are active agents or prisoners. Therefore, the system can change to the global minimum seeking consensus as legitimacy increases. The point of maximum energy is generally unstable for low values of police concentration because it can fall to the local or global minimum with variations in legitimacy. For high police concentrations, the maximum remains constant at low legitimacy, and as legitimacy increases, it reaches the minimum rapidly. Finally, the global energy minimum is stable for all cases because the system reaches a consensus. However, the increase of police officers' concentration makes its value change. Based on the results obtained, we can still interpret the system based on the principle of minimum grievance because the system tends towards a global minimum as legitimacy increases. However, the energy is insufficient to identify the system's prevailing state and the effects of police concentration on the dynamics.

Since the system now considers three possible states for agents and police officers, we must complement the analysis with the concentration of agents. With these two macroscopic quantities together, we can identify the most relevant function of the police based on the vision and the type of neighborhood. When considering vision one, the capture of active agents is the most relevant police function to dissuade a protest. In the Moore neighborhood with low police concentration, the system requires high legitimacy values for active officers are not predominant. As the concentration of police officers increases, we observe a significant decrease in the activity of active agents, so the system needs a lower value of legitimacy to change the active majority. In the random neighborhood, police officers increase the capture of active agents, allowing them not to be a dominant state in the system for low values of legitimacy. In addition, as the concentration of police officers increases, the activity of active agents rapidly decreases, increasing the concentration of jailed agents. Nevertheless, high values of legitimacy are still needed for passive agents to be predominant, and subsequently, the system reaches consensus. 

When the system has vision seven, we observe that there are still active agents with a low concentration of police officers, but they are not predominant. Therefore, the system requires increasing legitimacy to reach the predominance of passive agents. As the concentration of police officers increases, the activity of active agents disappears. As for legitimacy increases, the passive agents increase, and the jailed agents decrease linearly. Hence, in these cases, the relevant role of police officers is to prevent the appearance of a protest by preventing the officers from becoming active.

Although the police activity considerably reduces the active agents' activity in both cases, the permanent presence of jailed agents indicates a grievance in the population and an increase in the system's energy. Hence, we can conclude that the system will not be utterly stable if there is an internal grievance. This affirmation indicates that dissuading a protest by capturing active agents is ineffective in reducing grievance. Nevertheless, our results show that as the concentration of police officers in the system increases, the system needs a lower value of legitimacy to reach consensus or minimum energy. Observing the minimum obtained, we note that it is not an absolute minimum and depends directly on the concentration of police officers in the system. Therefore, the greater the concentration of police officers, the value of the minimum energy increases. In other words, dissuading protests to facilitate or maintain consensus has an energy cost for the system proportional to the number of police officers or the amount of force used.

We do not observe any significant variation of the system's behavior without police officers in terms of the variations in lattice size, the vision of the agents, or the type of neighborhood. So, we confirm that the dynamics only depend on the initial random conditions, such as the positioning of agents in the lattice and the definition of local parameters. Therefore, we conjecture that the system's dynamics will be the same regardless of the topology or dimension of the grid. In addition, the use of the movement rule will not generate state changes in the agents because the defined rules do not depend on the neighborhood in the system without police. In this context, the global quantities introduced and the phase diagram present new elements to analyze the model's dynamics and make new interpretations to understand the dynamics of social protests.

In the results obtained for the system with police officers, we did not notice significant variations of the dynamics with the variation of the size of the system. However, the dynamics dependent directly on the vision and the neighborhood product of the definition of the agents' rules. The police officers' activity increased significantly with the elections of a random neighborhood and increased vision. The global quantities introduced allowed us to identify the most relevant role and effect of the police officers in the system. It is essential to mention that the size of the system determines the maximum possible vision of the agents. Hence, for visions close to the maximum possible, we conjecture that the results will be similar to those shown for vision seven.

On the other hand, considering the movement of the agents and other topologies or lattice dimensions, we expect considerable changes in the system's dynamics with police officers. When Epstein presented this model in two dimensions, he reported very particular dynamics for specific parameters, such as the punctuated equilibrium phenomenon. Therefore, our results can serve future research as a first approximation to characterize the results reported by Epstein or even find new states or phenomena not yet reported in the current literature.

It is essential to note that we base our interpretations on the numerical results obtained from this simplified model. However, in the real world, the dynamics of social protests are more complex and involve many other factors. Nevertheless, studying this model from the perspective of sociophysics can yield new elements that allow us to address the complexities of the dynamics of social protest.

\begin{acknowledgments}
I.O. thanks ANID (Chile) for financial support through Doctoral Scholarship [BECAS-ANID/Doctorado Nacional/2019-21191380]. F.U. thanks FONDECYT (ANID-Chile) for financial support through Postdoctoral N$^\circ$ $3180227$. We acknowledge anonymous referees for their recommendations and constructive comments. 
\end{acknowledgments}

\bibliography{References}

\end{document}